\documentclass{aa}  
\usepackage{graphicx}
\usepackage{txfonts}
\usepackage{makecell}
\usepackage{chemfig} 
\usepackage{natbib}
\bibpunct{(}{)}{;}{a}{}{,} 
\usepackage[colorlinks=true, citecolor=blue, linkcolor=blue, urlcolor=blue]{hyperref}
\usepackage{stfloats}
\usepackage{lineno}
\usepackage{placeins}

\begin{document} 

    \title{When Two Worlds Collide: How Collisions Between Silicate and Carbonaceous Nanograins can Drive Complex Chemistry}
    
    \author{
    Alexandros Kyriazis\inst{1,2}
    \and
    Albert Rimola\inst{1,3}\thanks{Corresponding author: albert.rimola@uab.cat}
    \and
    Stefan T. Bromley\inst{2,4}\thanks{Corresponding author: s.bromley@ub.edu}
    }
    
    \institute{
    Departament de Química, Universitat Autònoma de Barcelona,
    08193 Bellaterra, Catalonia, Spain
    \and
    Departament de Química Física, Institut de Química Teòrica i Computacional (IQTC),
    Universitat de Barcelona, Barcelona, Spain
    \and
    Accademia delle Scienze di Torino,
    Via Maria Vittoria 3, 10123 Torino, Italy
    \and
    Institució Catalana de Recerca i Estudis Avançats (ICREA),
    Barcelona, Spain
    }
    
   \titlerunning{Collisions between silicate and carbonaceous nanograins}
    \authorrunning{Kyriazis et al.}


  \abstract
    {Observations have revealed the presence of silicate and carbonaceous material in interstellar grains, however, these components are generally assumed to belong to distinct dust populations. While some dust models do incorporate mixed silicate-carbonaceous grains, there is little evidence to support grain mixing mechanisms. In this work, we use atomistic simulations to investigate collisions between silicate and carbonaceous nanograins at velocities representative of a range of astrophysical environments.}
    {Our overall objective is to determine how collision velocity governs the interactions between silicate and carbonaceous nanograins. We aim to: i) identify collisional regimes capable of producing mixed silicate-carbonaceous grains and/or chemically complex molecular species; and ii) quantify fragmentation threshold velocities related to dust destruction.}
    {We performed molecular dynamics simulations employing a machine-learning force field to model head-on collisions between silicate and carbonaceous nanograins of comparable masses. Collision velocities span 1 to 11 km s$^{-1}$. For all collisions, we tracked the extent of grain-grain mixing and the formation of molecular fragments.}
    {We identified four velocity regimes: 1) $\lesssim$ 1.5 km s$^{-1}$, where grains bounce off one another; 2) $\approx$ 1.5-3.5 km s$^{-1}$, where sticking between the grains starts to occur; 3) $\approx$ 3.5-7.5 km s$^{-1}$, where grains tend to aggregate and form inter-grain chemical bonds, yielding stable mixed grains; and 4) $\gtrsim$ 7.5 km s$^{-1}$, where fragmentation dominates. The latter regime produces CO as the main product, along with hydrocarbons, complex organic molecules, molecular silicates, and mixed carbonaceous-silicate clusters. The fragmentation threshold velocity for these collisions is approximately 7.5 km s$^{-1}$.}
    {We show that collision velocities govern both the physical and chemical outcomes of silicate-carbonaceous nanograin interactions. In the fragmentation regime, collisions provide a viable pathway for generating mixed grains and a wide range of  molecular species, many of which have been observationally detected. Here, we provide a simple credible mechanism linking the physics of grain processing with observed complex interstellar chemistry.}

   
   \maketitle
   \nolinenumbers
\section{Introduction}

Dust grains make up only about one percent of the mass in the interstellar medium (ISM), yet they play a pivotal role in its chemistry and physics \citep{van_dishoeck_astrochemistry_2014}. These microscopic solid particles serve critical functions that include shielding ultraviolet radiation, catalysing chemical reactions, mediating energy transfer, and providing surfaces where atoms and molecules can adsorb and interact \citep{van_dishoeck_dust_processes,bromley_challenges_2014,potapov_dust_processes,pareras_single_atom_2023,jubert_H2_formation}. Such grain-driven processes are fundamental to producing much of the molecular inventory observed in the ISM.

Two primary classes of dust grains are observed: silicate grains (mainly composed of O, Mg, Si, and Fe) and carbonaceous grains (mainly composed of C and H). This chemical separation can be traced back to grain formation in the outflows of asymptotic giant branch (AGB) stars. In these circumstellar environments, the carbon to oxygen (C/O) ratio dictates the chemical composition of the forming dust. Carbon and oxygen first combine to create carbon monoxide molecules, which are highly stable and resist further reactions \citep{henning_cosmic_2010,pelaez_plasma_2018,gobrecht_molecules_2019}. When C/O<1, there is excess oxygen that readily reacts with silicon atoms to make SiO molecules; together with magnesium, this forms the basis for the nucleation of silicates \citep{bromley_silicate_nucleation}. Conversely, when C/O>1, the excess carbon drives the formation of carbonaceous dust \citep{c_o_1_agbs}. These conditions naturally separate the two dust families at formation. During circumstellar dust formation, grain mixing may rarely arise in binary systems where one star forms silicate dust and the other carbonaceous dust, which allows the two populations to coexist and potentially interact \citep{molster_iras_2001}.

After formation, dust is injected into the diffuse ISM where it is processed by supernova-driven shock waves. As these shocks propagate, they generate relatively large grain velocities producing frequent grain-to-grain collisions \citep{Tielens1994,Jones1994}. Collisional shattering is the dominant mechanism that efficiently redistributes dust mass from a minority population of large grains to a highly numerous population of progressively smaller grains and ultimately to nano-sized particles \citep{Jones1996,Bocchio2014}. Observations of non-radiative shocks confirm their capacity to induce significant dust destruction \citep{Zhu2019}. Shock models further suggest that the return of the smallest grains to the gas phase through complete destruction depends on the evolving shock structure and on the local interstellar conditions \citep{Slavin2015,Kirchschlager2022}.

Because ISM dust originates largely from both oxygen-rich and carbon-rich AGB outflows, many collisions in shocks are expected to involve both silicate and carbonaceous grains. The terminal stages of such collisional cascades are expected to produce nanograins and gas-phase species with compositions that reflect both silicate and carbonaceous precursors. Dust-evolution frameworks also predict that such cross-material interactions should occur in the ISM \citep{jones_evolution_2013,jones_global_2017,ysard_themis_2024}. Although shocks dominate the most destructive processing, a great deal of ISM dust also resides for long periods in more quiescent regions, such as translucent clouds and the warm inner zones of protoplanetary discs. In these more protected environments, grains remain bare or only lightly mantled and collisions are expected to be far less destructive than those driven by shocks. However, the available evidence indicates that the two dust families usually persist as separate populations throughout the ISM and no confirmed observations of mixed grains exist.

Laboratory studies of low-temperature condensation under astrophysically relevant conditions show that silicate and carbonaceous materials tend to nucleate and grow independently, rather than forming composite particles \citep{draine_interstellar_2009,rouille_separate_2020}. Astronomical observations show that C-H stretching bands associated with carbonaceous dust are essentially unpolarised, while the Si-O stretching features of silicates display clear polarisation. This provides strong evidence that the two dust families do not share the same physical environment \citep{chiar_spectropolarimetry_2006}. These findings also raise an important question in dust evolution regarding what happens when silicate and carbonaceous grains interact.

In this study, we modelled collisions between representative silicate and carbonaceous nanograins in the pre-destruction stage of their processing for a range of astrophysically relevant velocities. Our aim is to determine the threshold velocities that permit mixing, the velocities that cause full destruction, and the nature of the  post-collisional solid fragments and molecular species. Finally, we assess the potential role of such interactions with respect to the chemical evolution of the ISM.

\section{Methodology}

\begin{figure}
   \centering
   \includegraphics[width=0.90\hsize]{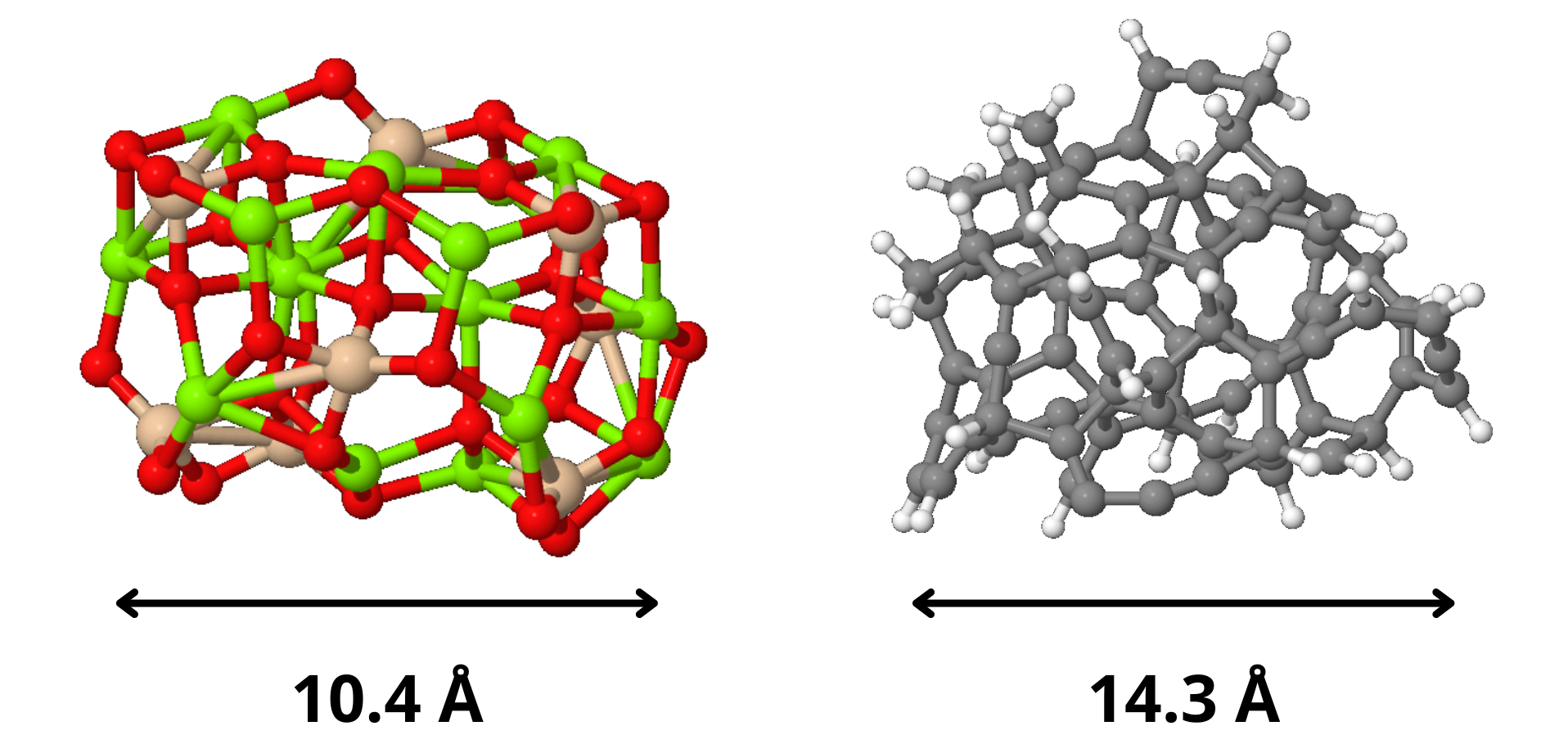}
      \caption{Nanograin models used in the collisions are an amorphous, magnesium-rich silicate with a forsterite stoichiometry and a $(\mathrm{Mg}_2\mathrm{SiO}_4)_8$ composition (left), and a hydrogenated amorphous carbon (HAC) nanograin with a $\mathrm{C}_{90}\mathrm{H}_{40}$ composition (right).    
              }
      \label{Structures}
   \end{figure}

\begin{figure}
   \centering
   \includegraphics[width=\hsize]{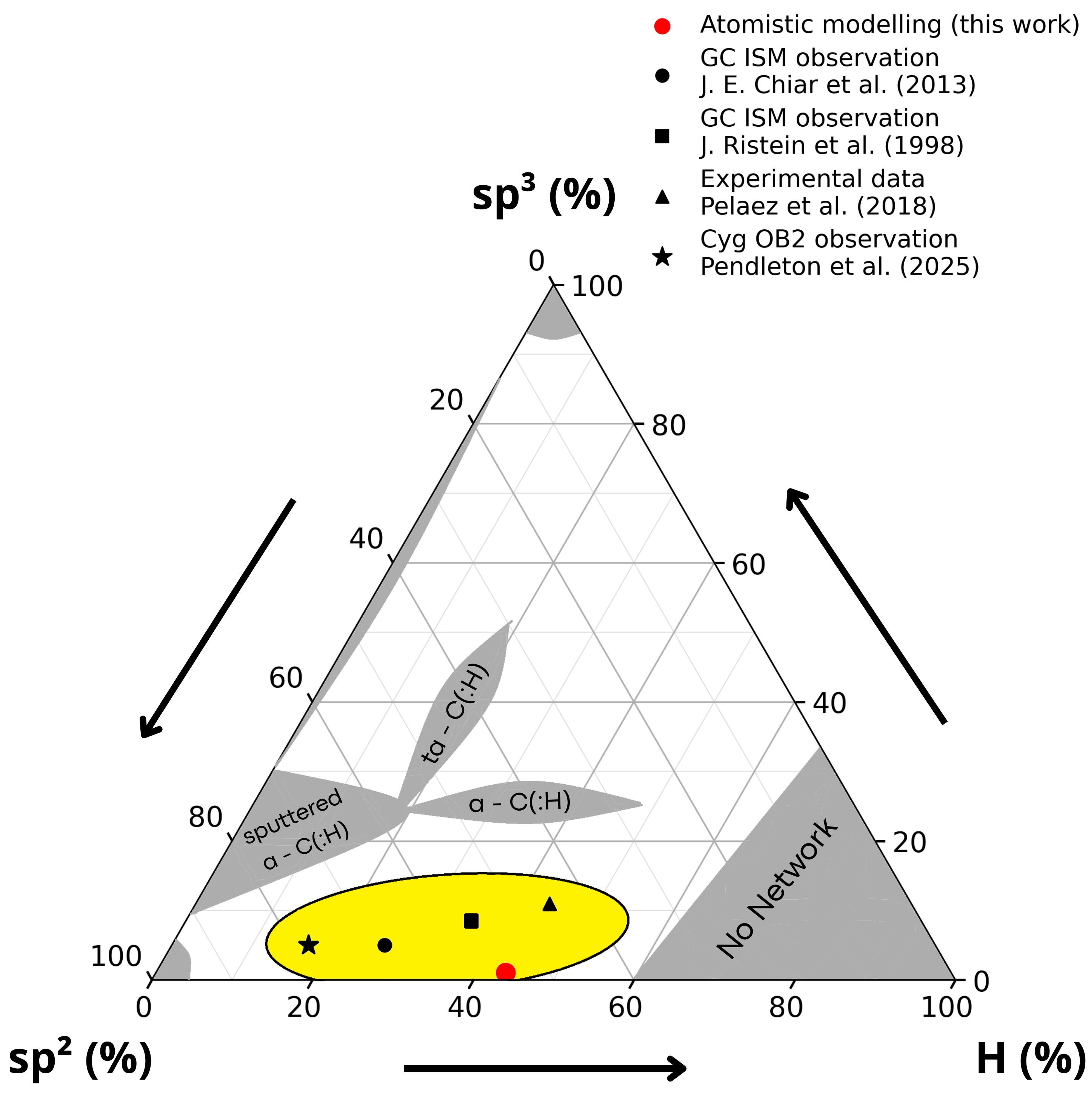}
      \caption{Ternary phase diagram illustrating the relative content of hydrogen, sp$^2$-, and sp$^3$-hybridised carbon in hydrocarbon materials. The labelled regions follow those in the detailed description provided in \cite{robertson_diamond-like_2002}. This figure is adapted from the ternary phase diagram presented by \cite{chiar_structure_2013}. In the yellow region, the red circle corresponds to the HAC structure utilised in our collision simulations. All other data points refer to IR observations of carbonaceous dust that have been spectral decomposed to obtain the three components: black cross \cite{chiar_structure_2013} and rhombus \cite{ristein_comparative_1998} represent observations toward the Galactic Centre; black star \cite{Pendleton_2025} shows an observation along a different line of sight (Cygnus OB2-12); black triangle \cite{pelaez_plasma_2018} represents experimental data.
              }
      \label{Ternary_diagram}
   \end{figure}

To simulate the collisions, we employed classical atomistic molecular dynamics (MD) simulations, which enable a detailed description of high-velocity collisions of nanograins at a reasonable computational cost. Our simulations model collisions between an amorphous magnesium-rich silicate nanoparticle with the stoichiometry of forsterite (Mg$_2$SiO$_4$)$_N$ and a hydrogenated amorphous carbonaceous (HAC) nanoparticle. The atomistic structure of the silicate nanograin was taken from a globally optimised model reported in previous work \citep{Global_optimized_silicates} with a (Mg$_2$SiO$_4$)$_8$ composition. The HAC nanograin model was generated through a two-step procedure using the AIREBO-M classical reactive potential \citep{airebo-m}. First, we performed a liquid quench on a 5000 carbon atom simulation cell following the approach in \cite{li_effect_2013,ranganathan_generation_2017}. Second, we extracted a discrete amorphous carbon nanoparticle from the obtained periodic glassy structure and hydrogenated it using the NAGGS code \citep{NAGGS}. Here, successive hydrogen atoms were stochastically collided with the particle until the desired hydrogenation was achieved. This procedure was tuned to provide a carbonaceous nanograin with a C$_{90}$H$_{40}$ composition, which has a very similar mass to that of the nanosilicate grain. Both the silicate and carbonaceous nanograins have diameters close to 1 nm (see Figure~\ref{Structures}). This small sizes of these grains represent the very latter stages of dust processing before total destruction. We note that the HAC nanograin contains proportions of \textit{sp}$^3$ carbon, \textit{sp}$^2$ carbon, and hydrogen that are are in good agreement with experimental and observational constraints. The corresponding data is shown in the ternary phase diagram in Figure~\ref{Ternary_diagram}. 

Key parameters that influence grain collision outcomes include relative grain sizes, collision angle, chemical composition, and collision velocity. In this study, we fixed the first three parameters and focused our investigation on the dependence of the outcomes on velocity. The angle of collision was set to zero, ensuring head-on collisions. The same principles were applied consistently throughout the entire study. The use of nanoparticles of the same mass also helped to eliminate net momentum transfer during the collision. As such, the  kinetic energy of the collision is channelled into deformation, fragmentation and chemical interactions rather than into post-collision net motions of the nanoparticles. 

To implement this setup, we first separated the two particles such that their surface-to-surface distance was $20\text{ \AA}$. The line connecting the centre of mass of the two particles defines a unique axis of approach. We then assigned initial velocities, so that the particles had equal and opposite momenta along the collisional axis so the motion was purely collinear and head on. The selected velocity magnitudes were 0.8, 1.8, 2.5, 4.0, 5.6, 6.9, 8.0, 8.9, 9.8, 10.6, and 11.3 km s$^{-1}$, which correspond to per atom kinetic energies of 0.01, 0.05, 0.1, 0.25, 0.5, 0.75, 1.0, 1.25, 1.5, 1.75, and 2.0 eV. To avoid bias caused by repeatedly colliding the same local surface regions of each particle, we generated ten random orientations for each particle through independent rotations around their centre of mass. As the internal atomistic structure of the grains is amorphous, rotational sampling also allows us to probe a fuller range of possible internal structural responses. The same set of rotational orientations was used for each velocity to allow for a fair comparison of different velocity regimes. The combination of rotational and velocity sampling produced a total of one hundred and ten independent simulated collisions. This large number of collisions also improved the statistical reliability of our results. 

Our computational modelling setup utilised a custom MD code based on the Atomic Simulation Environment (ASE), operating under the microcanonical (NVE) ensemble with a velocity Verlet integrator. As is standard in atomistic modelling, all processes were modelled as adiabatic and proceeded on the electronic ground-state potential energy surface. We note that our collisions involve energies per atom that are significantly lower than those expected to be needed for producing electronically excited species. Thus, while we do not rule out the possibility of some small proportion of non-adiabatic effects (e.g. ionisation) in our highest velocity collisions, we feel that the omission of such effects in our simulations is unlikely to affect our core results. Simulations were run for 4.5 ps, ensuring that for the higher velocities, the collision event occurred within 0.5 ps, allowing at least 4 ps for the system to evolve. The most critical factors affecting the simulation accuracy are the MD timestep and the reliability of the interatomic potentials. High-velocity collisions require a small enough timestep to avoid overly large atomic displacements in the MD propagation that can result in inaccurate force calculations and non-physical dynamical trajectories. We evaluated timesteps of 0.1, 0.05, and 0.01 fs across all the atomic pairs (details provided in Appendix~\ref{sec:appendix_b}), ultimately selecting 0.01 fs to ensure MD stability and accuracy. Each simulated collision thus consisted of 450000 MD steps.

The choice of a suitable interatomic potential to describe the wide range of possible interactions between the H, C, O, Mg, and Si atoms during the collisions is also crucial. Here, we employed a machine-learning interatomic potential (MLIP) based on the Message passing Atomic Cluster Expansion (MACE) architecture \citep{batatia_design_2022,batatia_foundation_2024,batatia_cross_2025}. Specifically, we used the \href{https://huggingface.co/mace-foundations/mace-mh-1}{MACE-MH-1} multihead model, pre-trained on large datasets of accurate quantum chemical density functional theory (DFT) calculations of many chemical structures. Specifically, MACE-MH-1 is trained using the OMat24 dataset (>100 million inorganic crystal structures) and fine-tuned on 1\% of the OMol25 dataset (>100 million organic molecules) \citep{barroso-luque_open_2024,levine_open_2025}. While the OMat24 pre-training provides extensive coverage of relaxed inorganic solid-state crystals, our simulations involved collisions of finite nanograins with non-crystalline structures and high surface-to-bulk ratios. MLIPs that are mainly trained on relaxed minimum energy structures tend to underestimate interatomic forces for structures away from equilibrium \citep{deng_systematic_2025}. This issue could affect the accuracy of our collisional simulations that inherently involve reactive intermediates, bond breaking and formation, and non-equilibrium geometries. Fine-tuning of foundational MLIPs has been shown to substantially reduce energy and force errors in such chemically complex regimes \citep{fine_tuning}. Here, fine-tuning using the OMol25 dataset is particularly suited to our system, as it contains a large collection of finite systems of up to 350 atoms with a wide range of intra- and intermolecular interactions, reactive structures, non-equilibrium geometries, and variable charge and spin states. Both the nanograin models that we consider fall well within this size regime and all five elements relevant to our simulations (H, C, O, Mg, Si) are represented in this dataset. To provide further confirmation of the suitability of our MLIP, we validated it by comparing the calculated forces and energies with those obtained from accurate DFT calculations for a representative collisional trajectory (see details in Appendix~\ref{sec:appendix_a}).

After each simulation, we analysed the outcome of the collision using the KDTree algorithm \citep{kd_tree} to identify the nearest neighbours of each atom. This was followed by a breadth-first search (BFS) \citep{BFS_algorithm} using interatomic separations to group nearby atoms into a set of distinct bonded clusters. Here, atoms were grouped together if their separation did not exceed their typical equilibrium bond length by more than 20\%. At the end of each simulation, the fragments were classified as either HAC or silicate successors based on the respective parent grain the majority of atoms in the fragment had originated from. We also used Open Visualization Tool (OVITO) \citep{ovito} and \href{http://www.jmol.org/}{Jmol} to visualise the trajectories.

\section{Results}

The outcomes of grain-grain collisions are strongly influenced by the relative velocity involved \citep{papoular_collisions_2004,guillet_shocks_2011}. The collisional velocity range explored in our simulations corresponds to a range of astrophysical environments. The slowest collision velocities (<1 km s$^{-1}$) are potentially relevant for protoplanetary disks, where such interactions mark the initial stages of planetesimal growth \citep{dominik_physics_1997,liu_growth_2019,tominaga_coagulation_2021}. Higher velocity grain-grain collisions are found across various astrophysical environments, including turbulent regions of the ISM ($\approx$ 0.1--5 km s$^{-1}$) \citep{yan_dust_2004}, denser regions such as the low-velocity end of molecular outflows ($\approx$ 1--5 km s$^{-1}$) \citep{arce_molecular_2006, low_velocity_outflows_velocity_2008}, low C-type shock regions ($\approx$ 1--10 km s$^{-1}$) \citep{van_loo_sputtering_2013, C_type_shocks_velocity_2016, desimone_train_2022, dubois_galaxies_2024}, and cloud-cloud collisions ($\approx$ 1--40 km s$^{-1}$) \citep{cloud_cloud_collisions, desimone_train_2022, cloud_cloud_collision_velocities_2025}. Below, we analyse the velocity-dependent outcomes of our mixed HAC-silicate nanograin collisions from both a physical and chemical perspective.

\subsection{Physical analysis of collisions}

\begin{figure}
   \centering
   \includegraphics[width=0.95\hsize]{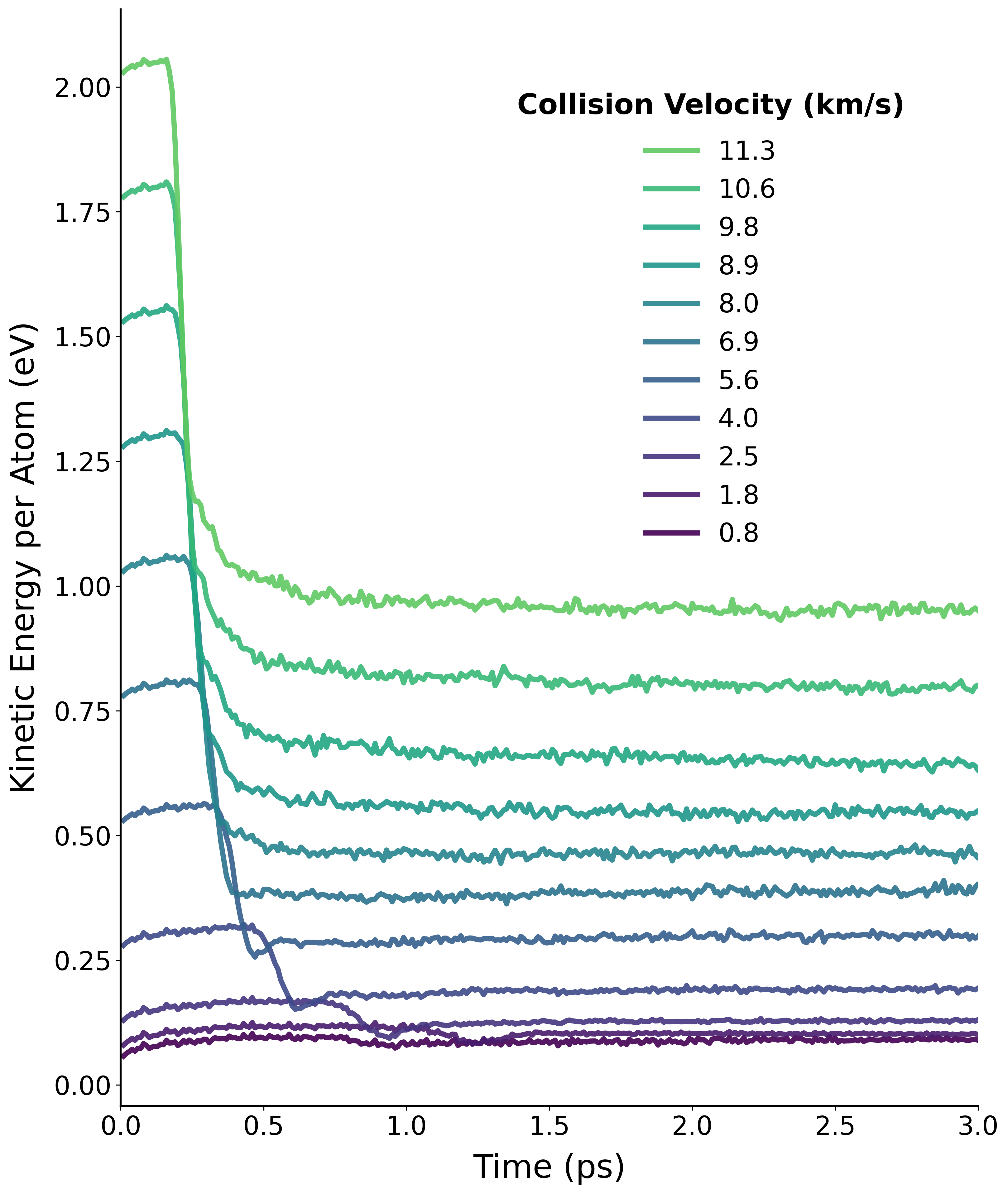}
      \caption{Time evolution of the kinetic energy of the colliding HAC-silicate nanograins for collision velocities ranging from 0.8 to 11.3 km s$^{-1}$. Each line corresponds to a different collision velocity.
              }
      \label{Kinetic_energies}
\end{figure}

\begin{figure*}
    \resizebox{\hsize}{!}{
        \includegraphics[width=0.80\hsize]{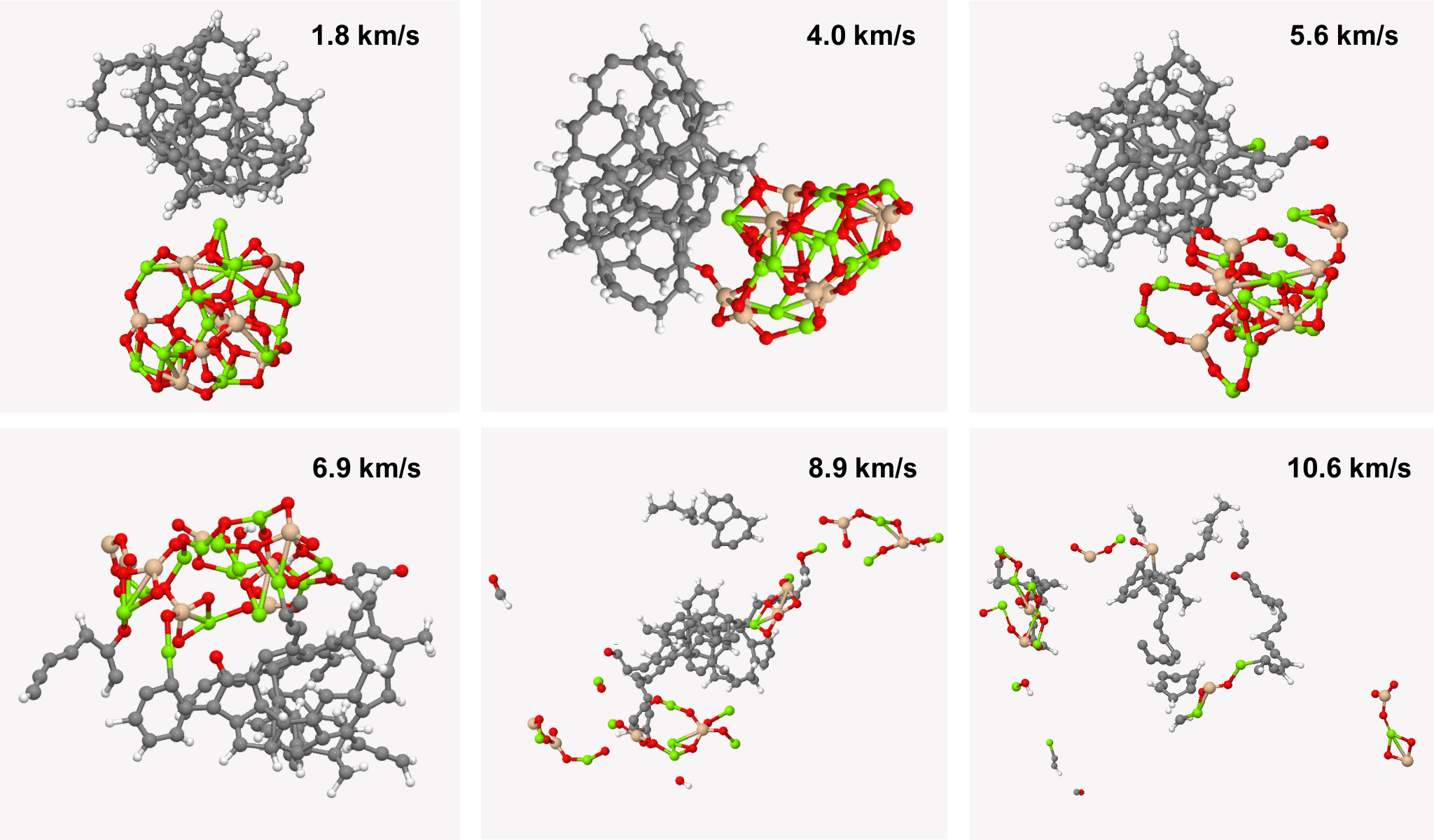}
    }
    \caption{Snapshots of the system 1 ps after the initial contact of the nanograins for different initial velocities. Atom colour key indicates hydrogen: white; carbon: dark gray; oxygen: red; silicon: beige; and magnesium: green.}
    \label{Photos after collision}
\end{figure*}

The outcome of the collisions between the silicate and HAC nanograins can be categorised based on key physical observables such as the energy dissipation, degree of fragmentation and mass distribution of the fragments. By analysing these parameters, we were able to gain insights into the physical mechanisms that govern the collision interaction between the two partners. Overall, in a microcanonical MD simulation, the total energy remains conserved. For any non-perfectly elastic collision, a significant proportion of the original kinetic energy of the colliding nanograins can be converted into potential energy of new interatomic interactions. In our simulations, the separated nanograin system starts with a relatively high translational kinetic energy and a low potential energy configuration. Upon collision, there is a sharp drop in kinetic energy, which is transferred into potential energy via new interatomic interactions (e.g. formation of new chemical bonds), while the remaining kinetic energy is redistributed into local heating, vibrational excitation, and rotational and translational motion of the resulting fragments. To quantify this local heating, we calculated the internal kinetic temperatures of the largest resulting silicon-based and carbon-based grains (see Table \ref{tab:temperatures}). Tracking this temperature is crucial for assessing the long-term stability of the grains, such as their likelihood of undergoing unimolecular decay. We computed these temperatures by subtracting the centre of mass velocity and the rotational motion of each species to isolate its internal vibrational degrees of freedom, and then applied the equipartition theorem. The results reveal a consistent physical trend: at the lowest velocities, the species exhibit moderate heating ($\approx$300–800 K), while at the highest velocities, these temperatures reach $\approx$2000–4000 K, peaking at $\sim$5000 K. We emphasise that these values represent the peak internal temperature attained immediately after the collision. This energy will be rapidly radiated away primarily via infrared (IR) photon emission from vibrationally excited modes.

In Figure~\ref{Kinetic_energies}, we show how the kinetic energy of the system changes during the collision simulations for the different considered collision velocities. Figure~\ref{Photos after collision} shows a snapshot of the resulting atomistic structure after 1 ps of a representative simulation for six different velocities in this considered range. Prior to the main collision event, a small increase in kinetic energy is observed in all simulations. This reflects the acceleration of the nanograins as short-range attractive forces convert potential energy into kinetic energy once the grains enter their mutual interaction range. For the three lowest velocities (0.8, 1.8 and 2.5 km s$^{-1}$), the loss of kinetic energy is very low (<0.1 eV/atom). In these cases, the grains deform only slightly upon colliding and bounce in an almost elastic fashion. For slightly higher collision velocities (4 - 5.6 km s$^{-1}$), the collisions become more inelastic and the fraction of the initial kinetic energy converted into chemical interactions rises to $\approx$50\%. This is the start of the regime where the grains deform or even start to break apart and lose their original structure as they interact upon aggregation. At the highest considered velocities (>8 km s$^{-1}$) we observe significant fragmentation and numerous associated chemical transformations. In this regime, increasing the velocity leads to an increase in the number of collisional fragments while the total number of interatomic interactions decreases. Although the structural outcome of the collisions sensitively depends on the collision velocity, the proportion of kinetic energy transformed into potential energy remains at around 50\% for all velocities above 4 km s$^{-1}$. This indicates that collisions progressively explore the complex potential energy landscape of possible chemical structures, enabling the formation of stable molecules beyond simple fragments of the original grains.

\begin{figure}
   \centering
   \includegraphics[width=\hsize]{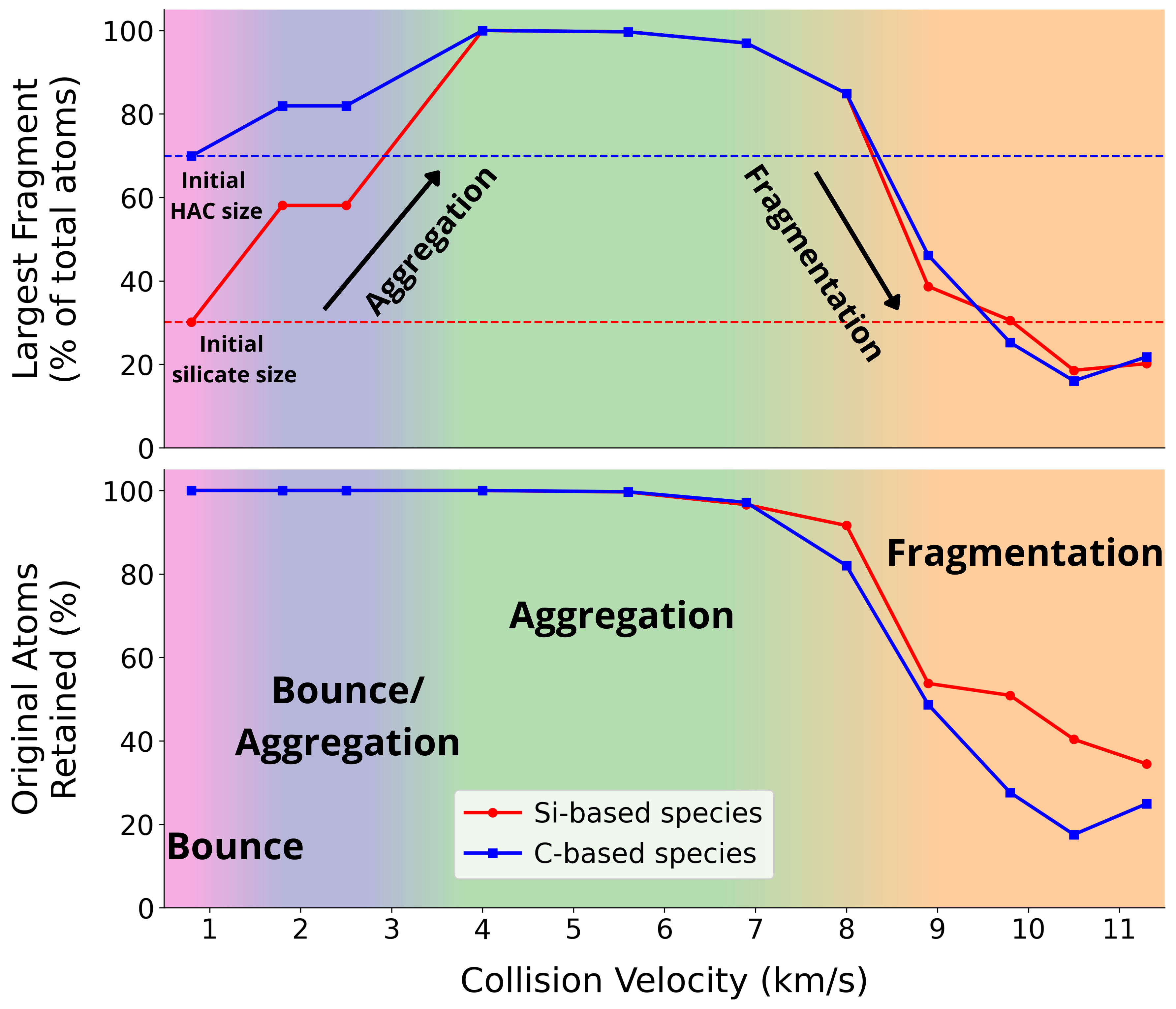}
      \caption{Average percentage of system atoms contained in the largest silicon-bearing species (red) and the largest carbon-bearing species (blue) as a function of collision velocity (top panel), and the percentage of original atoms retained in each grain (bottom panel). The dashed lines indicate the initial fraction of atoms belonging to each grain. The pink shaded region indicates the bounce regime, the blue region corresponds to bounce/aggregation, the green region shows the aggregation regime, and the orange region represents the fragmentation regime.
              }
      \label{Largest_cluster_analysis}
\end{figure}

In Figure~\ref{Largest_cluster_analysis}, we distinguish four different velocity regions: bounce ($\lesssim$ 1.5 km s$^{-1}$, pink region), bounce-and-aggregation ($\approx$ 1.5-3.5 km s$^{-1}$, blue region), aggregation ($\approx$ 3.5-7.5 km s$^{-1}$, green region), and fragmentation ($\gtrsim$ 7.5 km s$^{-1}$, orange region). The top panel of Figure~\ref{Largest_cluster_analysis} shows the percentage of the total atoms in the largest species as a function of the collision velocity. The bottom panel of Figure~\ref{Largest_cluster_analysis} shows the percentage of the atoms that are retained by each original nanograin after the collision. 

In the bounce regime, at the lowest considered velocity (0.8 km s$^{-1}$), the initial grains remain structurally intact as distinct separated species. Here, the largest species thus correspond to the sizes of the original grains. As the collisional velocity increases to between 2-3 km s$^{-1}$, the grains occasionally stick together. In this bounce-and-aggregation regime, the average size of the largest species increases depending on the sticking probability, which was found to be approximately 40\%. This kind of sticking behaviour is consistent with previous grain-grain collision studies \citep{jones_grain_1996, dominik_physics_1997, papoular_collisions_2004, nietiadi_collisions_2020}. However, these previous works differ fundamentally from the present study in terms of both methodology and scope. Jones et al. (1996) developed an analytical model for grain shattering based on continuum shock physics, while Dominik \& Tielens (1997) modelled grain aggregation by treating grains as elastic spheres interacting through macroscopic contact mechanics, without resolving individual atomic interactions. Papoular (2004) employed molecular dynamics to study grain-grain collisions, but considered carbonaceous grains only. This latter study also used semi-empirical interatomic potentials which, unlike MLIPs, are unlikely to be able to accurately capture the complex chemistry involved in grain-grain collisions. Similarly, Nietiadi et al. (2020) used a classical interatomic potential to simulate the collisions of bare amorphous carbon nanoparticles. The hydrogenated nature of our HAC grains passivates their surface, which explains the presence of a bouncing regime preceding sticking.

At higher velocities (3.5-7.5 km s$^{-1}$), the system transitions to the aggregation regime, where nearly all collisions result in a single combined species. As such, the largest species correspond to the total atom count. In this velocity range the collisional induced aggregation also leads to a large internal structural re-organisation of each grain. A detailed structural analysis of the resulting mixed grains is presented in Section 3.3. 

Above 7.5 km s$^{-1}$, the kinetic energy becomes sufficient to break-up the initial grain material into fragments. In this fragmentation regime, we observe a clear drop in the number of retained atoms after each collision, along with a corresponding sharp decrease in the size of the largest fragment. This effect is clearer for the HAC grain, which shows a notably higher loss of original atoms with increasing collisional velocity in this regime than the silicate grain. The lower silicate grain fragmentation rate is likely due to its stronger bonding in this inorganic refractory material and its more compact atomistic structure. Irrespective of the fragmentation rate, we identify that the velocity fragmentation threshold for both silicate and HAC nanograins is around 7.5 km s$^{-1}$, which for the case of the  silicates is in agreement with recent atomistic MD simulations of silicate nanograin fragmentation \citep{silicate_collisions_esmerian_2026}. We note that this critical velocity is higher than widely accepted values from previous grain collision studies \citep{jones_grain_1994,jones_grain_1996, hydrogenated_amorphous_carbon_parameters,guillet_shocks_2011, hirashita_evolution_2013, bocchio_re-evaluation_2014}. It is important to note that this threshold velocity is specific to our simulations involving mixed HAC-silicate nanograin collisions with fixed stoichiometry and size. A better estimate of the fragmentation threshold velocity for each grain family would require testing across a wider range of parameters (e.g. grain structure, composition, and size). 

Beyond 9.5 km s$^{-1}$, the largest fragment recovered from both silicate and HAC grains contains fewer atoms than the initial structure. At these sizes, further fragmentation becomes increasingly difficult due to the reduced number of atoms in the remaining fragments. The persistence of molecular-sized fragments shows that the grains are still not totally destroyed (i.e. reduced to their atomic constituents) in this velocity regime. 

\subsection{Chemical analysis of collisions}

We go on to examine the outcomes of our collision simulations from a chemical perspective. One aspect of particular interest is the point when the initial kinetic energy of the separate colliding nanograins is used for bond breaking and formation, enabling complex chemical transformations. Here, our focus is on the number and chemical composition of the species, the identification of specific molecules, and the atomic rearrangements within the system. 

\begin{figure}
   \centering
   \includegraphics[width=\hsize]{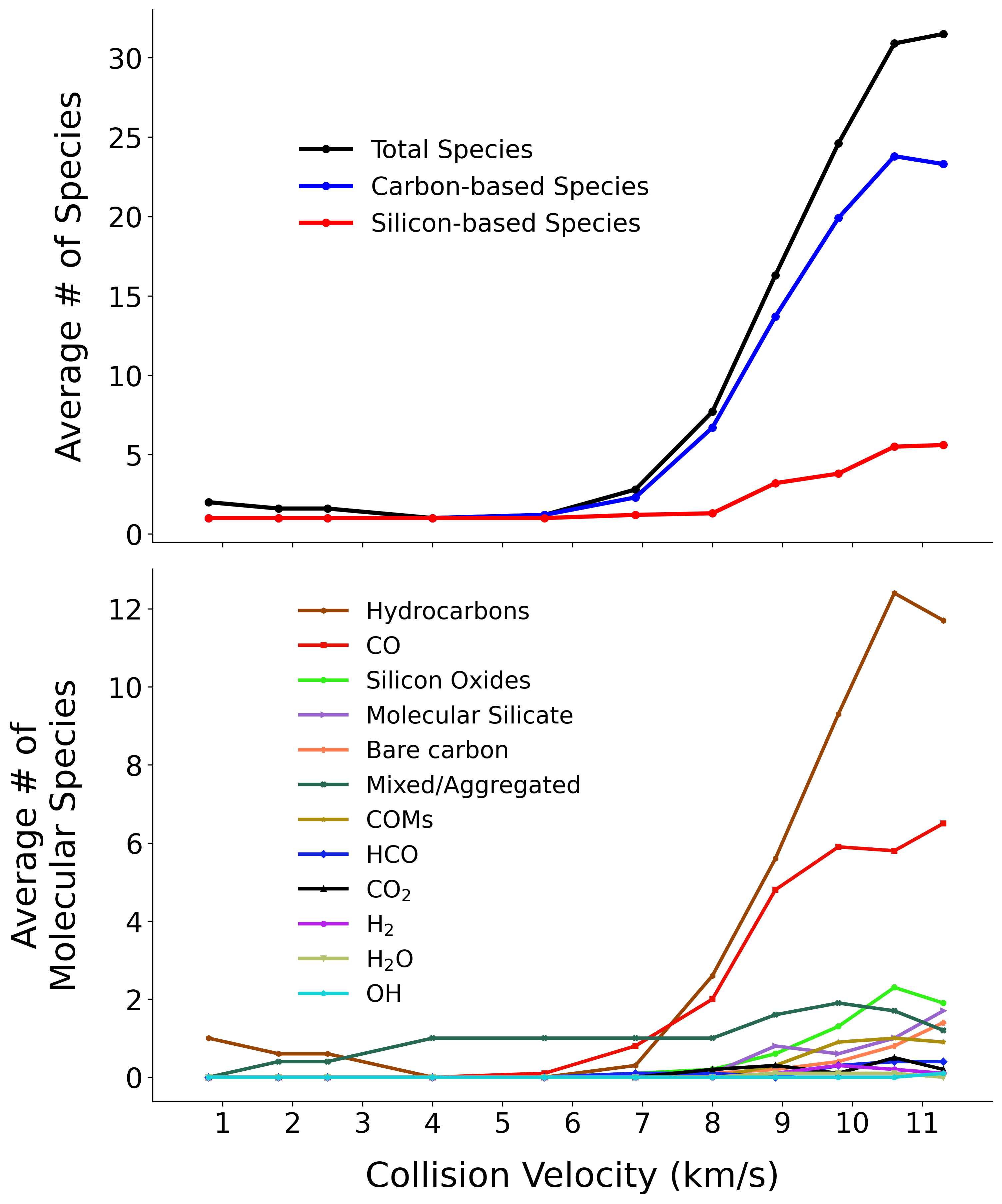}
      \caption{Top: Average number of total (black line), carbon-based (blue line), and silicon-based (red line) species formed per collision, plotted as a function of collision velocity. Bottom: Average number of selected chemical species present per collision as a function of collision velocity. The species are in order of abundance at the highest tested velocity of collision: hydrocarbons (brown), CO (red), silicon oxides (light green), molecular silicate (lilac), bare carbon (orange), mixed-aggregated species (dark grey), COMs (dark olive), HCO (blue), CO$_2$ (black), H$_2$ (magenta), H$_2$O (light olive), and OH (cyan).
              }
      \label{Average_cluster_count}
\end{figure}

In the upper panel of Figure~\ref{Average_cluster_count}, we show the average number of produced species per collision with respect to the collision velocity. In line with Figure~\ref{Largest_cluster_analysis}, we see that the total number of collision-induced species is very low up to a velocity of around 7 km s$^{-1}$, where it then starts to increase rapidly. We note that the number of species tends to saturate at the highest higher velocities considered. This likely results from the exhaustion of available atoms in the simulations. Broadly speaking, the species can be chemically categorised as being either carbon-rich or silicon-rich. This emergent separation likely arises from the higher energetic stability of the C-C and Si-O bonds with respect to Si-C bonds, tending to limit the formation of silicon carbide-based species. Of the two main classes of species, carbon-based species are found to be significantly more numerous than silicon-bearing species. 

To gain deeper insights into the chemical evolution of the system, we identify and quantify specific molecular species formed as a function of collision velocity. Understanding the nature of these molecules provides valuable information about the post-collision chemical environment and helps identify potential tracer species for observational studies. The bottom panel of Figure~\ref{Average_cluster_count} records the number of newly-formed species across the range of collision velocities. Here, the dominant carbon-based fraction of fragments in the high velocity collisions is mainly comprised of hydrocarbons and CO molecules. To understand the origin of these two types of carbon-based fragments, we considered two key processes. 

At sufficiently high collision velocities, both grains deform upon collision, which leads to a relatively high and intimate grain-grain surface contact area. This situation allows reactive carbon atoms (e.g. undercoordinated surface atoms) from the HAC grain to abstract oxygen atoms from the silicate. This favours the formation of highly stable and volatile CO molecules, which subsequently detach from the grains. This process is likely to be both thermodynamically and kinetically favoured under such high-energy conditions. The production of CO increases with increasing collision velocity throughout the fragmentation regime and peaks at 11.3 km s$^{-1}$ with an average of 6.5 molecules produced per collision. As CO is one of most abundant collisional products, it could potentially be used as a tracer, along with other fragment fingerprints, to detect regions where bare HAC-silicate grain interactions could occur. 

The most abundant molecular species formed in our HAC-silicate collisions in the fragmentation regime are hydrocarbons (i.e. species composed solely of carbon and hydrogen atoms). Unlike CO, the production of such species does not require intergrain chemistry, and can be rationalised via the physical fracturing of the HAC grain after high velocity collisions. The propensity of such a process is likely to be higher for the HAC nanograin due to the lower cohesive energy of typical covalent C-C and C-H bonds compared to the stronger more ionic bonding in the nanosilicate grain. In our simulations we find that hydrocarbon production increases rapidly upon entering the fragmentation regime, and peaks at 12.4 fragments per collision at 10.6 km s$^{-1}$.

After 8 km s$^{-1}$, when fragmentation becomes significant for both grains, other important classes of molecules begin to emerge, such as complex organic molecules (COMs). In this study, we define COMs as species composed of 6--30 atoms, containing carbon and at least another heavier atom (i.e. O, Si, and/or Mg), and exhibiting a higher carbonaceous character relative to silicate content. This definition slightly differs from the conventional COM definition (i.e. molecules with six or more atoms containing at least one carbon atom) as it helps filter out misclassifications. Without these upper atom-count and elemental constraints, large silicate-rich or pure hydrocarbon species might otherwise be incorrectly identified as COMs.

COMs begin to form after the fragmentation threshold, starting at 0.3 molecules per collision at 8.9 km s$^{-1}$. Their rate of formation also increases with increasing velocity, reaching around a molecule per collision at the highest considered velocities. This production route could add a significant contribution of COMs to the interstellar population. In addition to COMs, our collisions also produce key COM precursors. Notably, the HCO (formyl) radical, a key species in the formation of COMs \citep{enrique-romero_reactivity_2019,Enrique-Romero2022} is a common product once fragmentation begins and is produced as a rate of 0.4 molecules per collision at velocities above 9.8 km s$^{-1}$. These results suggest that the abundant reservoir of H, C, and O atoms. The high excess energy of the collisions can provide conditions for activating post-collisional chemical reactions between produced fragments.

Next, we quantified the formation of silicon oxide-based species (Si$_y$O$_x$), with particular focus on the silicon oxide (SiO) molecule. This common gas-phase species is often used as tracer of shocks in dense molecular clouds, in molecular outflows, and in cloud-cloud collisions \citep{cloud_cloud_collisions, desimone_train_2022, rojas-garcia_interstellar_2022, rybarczyk_shaken_2023}. According to our results, Si$_y$O$_x$ species lock up further oxygen and begin to form following the fragmentation of silicates, as expected \citep{Andreu_silicon_oxides_2024}. Once initiated, their abundance increases rapidly, reaching up to 2.3 molecules per collision at 10.6 km s$^{-1}$, becoming the third most prominent chemical species after hydrocarbons and CO.

The remainder of the initial silicate grain is redistributed into molecular silicates. Such species retain the three character atoms of the parent silicate grain (i.e. Mg, si, and O) and may also bear hydroxyl groups in cases where hydrogen transfer has occurred. Such species are of considerable astrophysical interest, as they are thought to constitute the basic building blocks of monomer-initiated silicate nucleation \citep{Joan_oxygen_rich_2025, Gelli_2025} and have been directly observed in cluster beam experiments \citep{Joan_molecular_silicate_beam_2022}. Interestingly, in our grain collisions the stoichiometry of the resulting molecular silicate species can differ from the forsteritic stoichiometry (Mg$_2$SiO$_4$) of the parent grain. For example, we find molecular silicates with enstatite stoichiometry (MgSiO$_3$) or even adopt more oxygen-poor compositions such as MgSiO$_2$. The most commonly formed molecular silicate species across our simulations are summarised in Table~\ref{table:species_summary}. In our simulations, the production of molecular silicates increases rapidly with collision velocity. Their detection begins at 8.0 km s$^{-1}$, with only 0.1 species per collision, and rises quickly, at 10.6 km s$^{-1}$ exceeding one molecular silicate per collision. This suggests that high-energy grain collisions could represent an efficient pathway for injecting silicate monomers into the gas phase, where they could potentially subsequently participate in nucleation and growth processes. In contrast, the formation rates of H$_2$O, CO$_2$, and H$_2$ remain below one molecule per collision across all studied velocities. The CO$_2$ formation reaches a maximum of 0.5 molecules per collision only at 10.6 km s$^{-1}$, where grain fragmentation is already significant. H$_2$ and H$_2$O formation appears particularly inefficient in these collisions, with a maximum of only 0.3 and 0.1 molecules per collision respectively.

\subsection{Mixed and aggregated HAC-silicate species}

Here, we examined the extent of the HAC-silicate mixing formed in our collisions. To do so, we identified the types of chemical bonds of the mixed species after each collision. To carry out the analysis, we extracted all mixed species, structures containing all five atom types (i.e. H, C, O, Mg, and Si), resulting from each collision and optimised their structures (i.e. relaxed them to their 0K geometries). The purpose of this optimisation was to approximate the thermal dissipation of the excess energy of the species well after the collisions. This procedure also helps to cleanly analyse characteristic bond length distributions, which would be statistically more noisy with more distorted, non-equilibrium structures from energetically excited species directly after a collision.

\begin{figure}
   \centering
   \includegraphics[width=0.95\hsize]{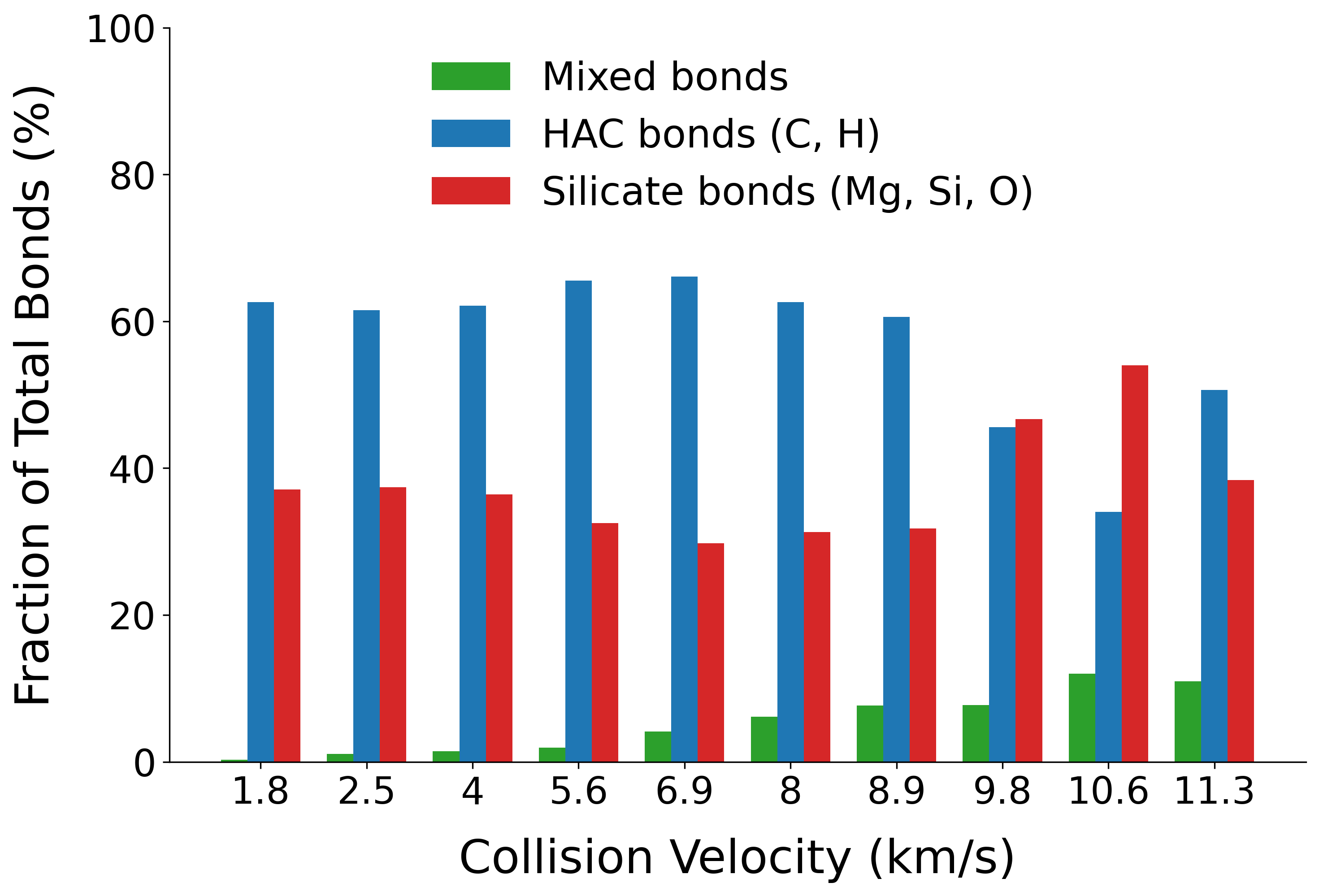}
      \caption{Fraction of bond types formed in the mixed/aggregated species as a function of collision velocity, categorised as HAC–HAC (blue), silicate–silicate (red), and mixed bonds (green).
      }
      \label{mixed_clusters_bond_count}
\end{figure}

For these post-optimised structures, we calculated the fraction of the selected bonds over the total number of bonds and classified them as intra- or inter-grain bonds. Inter-grain bonds refer to HAC (C-H, C-C) and silicate (Si-O, Mg-O, Si-Mg) bond types, which are present in the grains before the collisions. Intra-grain or mixed bonds are those formed from bonds where each atom comes from a different grain. This later selection of bonds (e.g. C-O, Mg-H) indicates when mixed nanograin chemistry has occurred. Figure~\ref{mixed_clusters_bond_count} shows the average fraction of bond types as a function of collision velocity. For each collision velocity, all formed mixed species were collected and the reported values were averaged over the number of mixed clusters identified at that velocity. Generally, HAC bonds tend to dominate, which is largely due to the HAC nanograin possessing a higher number of atoms than the silicate nanograin. At low velocities, the HAC and silicate components remain largely chemically separated, with very few mixed bonds. The mixed bonds formed in low velocity collisions are typically the result of interfacial interactions upon partial sticking of the grains, with the main contributions being attributed to C-O, Mg-C or O-H bonds. We note that a significant fraction of the O-H bonds correspond to silanol groups (Si-OH) formed on the silicate grain from transfer of hydrogen from the HAC grain. As the velocity increases and the system enters the aggregation regime, the fraction of silicate bonds decreases slightly while the fraction of mixed bonds increases. We note that this is the result of interfacial grain-grain interactions, rather than the liberation of CO molecules, which is mainly found at higher velocity collisions (see lower panel of Figure~\ref{Average_cluster_count}). For further increases of collisional velocity up to 10.6 km s$^{-1}$, the fraction of mixed bonds continues to increase. In this velocity range the C-O bonding fraction decreases, as CO is liberated as free molecules and the mixed bond fraction is contributed mainly by O-H and C-Mg bonds. This indicates that fragments produced at this velocity range are the most chemically mixed. The formation of C-Mg bonds likely arises because magnesium is more weakly bound within the silicate structure and is therefore more prone to forming new bonds during high-energy collisions, particularly with reactive carbon species. Such bonds have recently attracted growing attention both computationally and experimentally, in efforts to explain the presence of Mg-bearing carbonaceous species observed in the circumstellar envelope of IRC+10216 and in meteoritic samples \citep{cernicharo_Mg_species,Metal-organics_in_meteorites,Mg_C6_H2_isomers,MgC3H_computational_study}.

Finally, we consider how the collisional kinetic energy leads to the chemical processing of the nanograins. To quantify collision-induced changes in the atomistic structure of each nanograin, we computed the atomic pair count $N_{ab}(r)$ as a function of interatomic distance for a selection of bond types,

\begin{equation}
    N_{ab}(r) = \sum_{i \in a} \sum_{j \in b} 
    \mathbf{1}\!\left[r \leq r_{ij} < r + \Delta r\right],
\end{equation}

\noindent
where $r_{ij}$ is the distance between atom $i$ of type $a$ and atom $j$ of type $b$, $\Delta r$ is the bin width, and  $\mathbf{1}[\cdot]$ is the indicator function. In addition, $N_{ab}(r)$ is not divided by the number of atoms or system volume and therefore it serves as a direct measurement of the total number of pairs at a given separation for each collision. 

\begin{figure*}
    \resizebox{\hsize}{!}{
        \includegraphics[width=0.95\hsize]{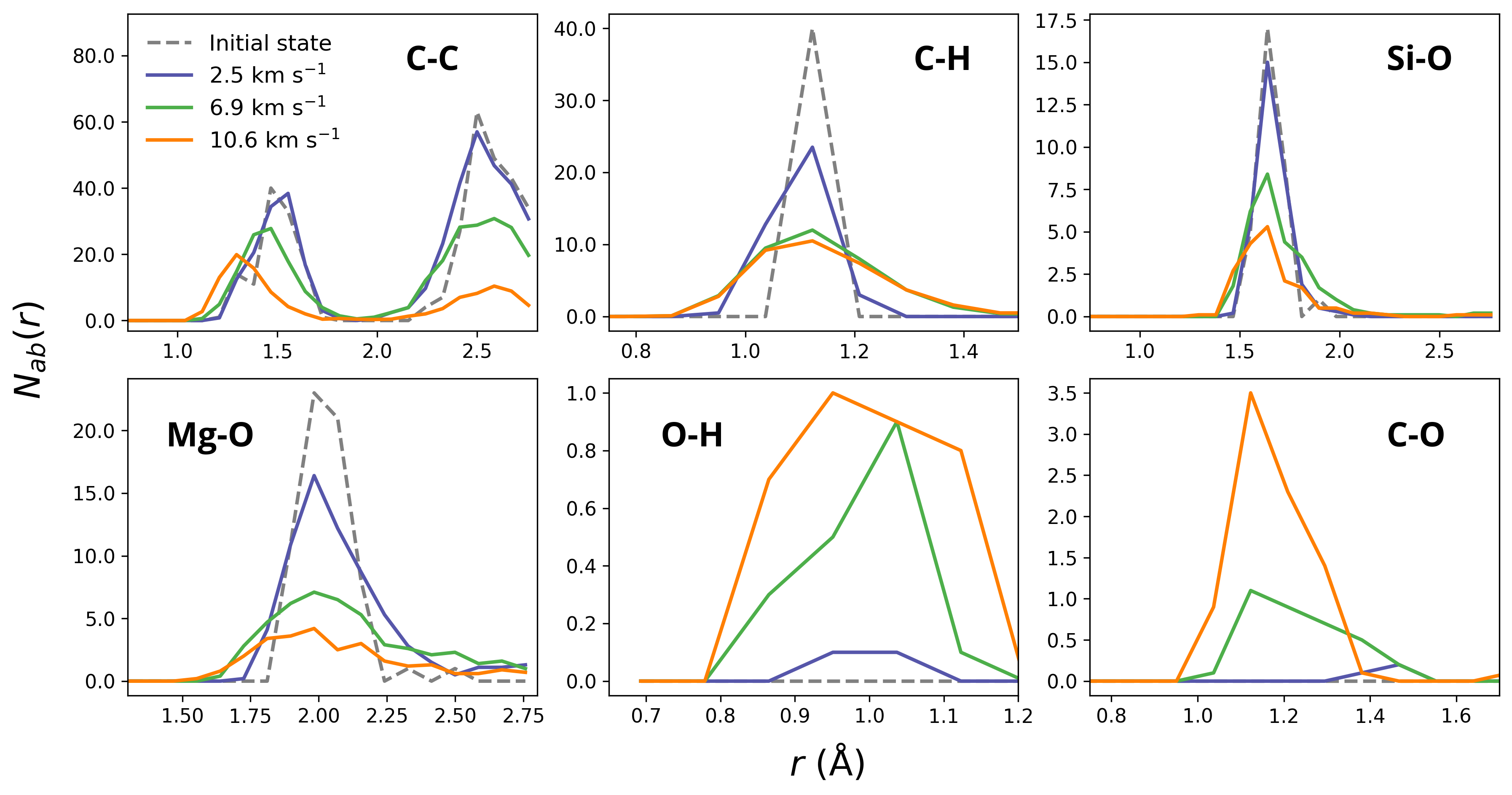}
    }
    \caption{Absolute partial pair counts $N_{ab}(r)$ for selected atom pairs as a function of interatomic distance,  $r$, computed at the final frame of simulations at three representative collision velocities: 2.5 km s$^{-1}$ (bounce-and-aggregation regime, blue), 6.9 km s$^{-1}$ (aggregation regime: green), and 10.6 km s$^{-1}$ (fragmentation regime: orange). The grey dashed line shows the initial pre-collision state, serving as a reference baseline. Each panel corresponds to a single pair type: C-C, C-H, Si-O, Mg-O, C-O, and O-H. $N_{ab}(r)$ counts the total number of $a$-$b$ pairs within each distance bin $[r,\, r+\Delta r)$}
    \label{Partial_rdfs}
\end{figure*}

Figure~\ref{Partial_rdfs} shows $N_{ab}(r)$ for six characteristic bonding pairs, each displayed as a function of velocity for three collision regimes (see Figure~\ref{Largest_cluster_analysis}), together with the initial pre-collision distributions shown as a grey dashed line for reference. The results for $N_{CC}(r)$ show two well-separated peaks corresponding to the first and second neighbour peaks near 1.5 and 2.5 $\AA$, respectively. As the velocity increases, the total number of C-C pairs decreases substantially, reflecting the fragmentation of the HAC nanograin. Additionally, the second-neighbour peak decreases more rapidly than the first, indicating that fragments are becoming progressively smaller. The C–C bond lengths also shift toward lower values with increasing collision velocity. To analyse this tendency, we computed the average C–C coordination number (CN) separately for short bonds ($r < 1.35$ \AA, indicative of sp-hybridised triple bonds as in polyyne-type species) and longer bonds ($1.35 \leq r < 1.85$ \AA, associated with sp$^2$ and sp$^3$ bonding environments). In the initial state, the bonded carbon network is dominated by long bonds (CN$_{long}$ = 2.27), with relatively few short bonds (CN$_{short}$ = 0.33). This distribution remains largely unchanged at 2.5 km s$^{-1}$. At 6.9 km s$^{-1}$, a structural shift becomes apparent: the contribution from long bonds decreases (CN$_{long}$ = 1.87), while that coming from short bonds increases (CN$_{short}$ = 0.45). In the fragmentation regime at 10.6 km s$^{-1}$, a marked inversion is observed: CN$_{long}$ drops to 0.69, while CN$_{short}$ rises to 0.79, and the total C–C coordination decreases to 1.48. This pronounced shift provides quantitative evidence of enhanced formation of linear carbon chains at high collision velocities, consistent with the increased production of small polyyne-type species observed in Table~\ref{table:species_summary}. Polyynes are among the most abundant organic molecules detected in the interstellar medium and their formation via high-velocity grain collisions might therefore represent a significant source of these species in shocked regions. The analysis of C-H pairs shows a single characteristic peak around 1.1 $\AA$ that decreases in intensity with increasing collision velocity. This reduction is linked not only to the de-hydrogenation of carbon atoms but also to a broadening of the C-H length distribution, caused by stronger vibrational motions at higher velocities. The single peaks associated with Mg-O and Si-O bonds show a systematic reduction in terms of the pair count with increasing velocity, consistent with the disruption of the silicate network. Increasing the collision velocity also leads to an increase in the number of O-H pairs at around $\sim$1.0 $\AA$, indicating the formation of silanol (Si-OH) groups. Additionally, a C-O coordination pair begins to appear, which can be attributed both to the formation of CO molecules and to interactions within the mixed aggregated grains. In the latter scenario, any so-called under-coordinated carbon atoms in the HAC interact with oxygen atoms in the nanosilicate matrix, thereby creating bound carbonyl groups (i.e. =C=O). We note that this process could be a contributing factor in oxygen depletion \citep{jones_depletion_oxygen_2019}. 

\section{Discussion and astrophysical implications}

The aggregation of HAC and silicate grains through grain-grain collisions is not a straightforward outcome, as it depends critically on the relative collision velocity. Collisions lead to sticking and subsequent aggregate growth only within a limited velocity range above the bouncing and below the fragmentation regimes. This narrow growth window is further constrained by the compositional mismatch between HAC and silicate materials, which differ in their atomic structure, bonding characteristics, surface energies, and mechanical properties. Therefore, in dynamically active environments characterised by a broad distribution of collision velocities, the efficient and net growth of HAC-silicate aggregation is expected to occur only in a relatively small subset of collision events. However, mixed and aggregated species can still be produced in the fragmentation regime, although with progressively smaller characteristic sizes as the collision velocity increases.

The threshold velocity at which grains undergo fragmentation has long been a key parameter in models of dust evolution, but recent atomistic simulations do suggest that the canonical values may have been underestimated. Molecular dynamics simulations of collisions between amorphous silica grains with radii in the range 6.5-50 $\AA$, at velocities between 0.1 and 20 km s$^{-1}$, reveal grain shattering thresholds at approximately 6 km s$^{-1}$, roughly a factor of two higher than the canonical value that is widely used in many astronomical models \citep{silicate_collisions_esmerian_2026}. Our results are in line with this observation and we further find evidence that the HAC shattering threshold may also be underestimated. This revision has direct implications for grain growth in shocked environments, as collisions in the range previously assumed to cause disruption may instead result in aggregation, shifting the predicted size distribution toward larger grains. Analogous surface effects have been reported for silica, where hydroxylation or ice coverage reduces surface energy and measurably alters collision outcomes \citep{Hydroxylated_silica_adhesion, nietiadi_hydroxylated_silica, nietiadi_ice_covered_silica}. Although silica is compositionally distinct from the silicate grains considered here, these findings underscore a broader sensitivity of collision dynamics to surface chemical composition and processing.

The hydroxylation of silicate grain surfaces through the chemisorption of hydrogen atoms, forming surface Si-OH silanol groups, is now well established theoretically, experimentally \citep{kerkeni_how_2017}. Hydroxylated silicates have also been observed via the 3.2 $\mu$m absorption band on comet 67P/Churyumov–Gerasimenko \citep{mennella_hydroxylated_2020}. Silanol groups have also been predicted to enhance the catalytic power of silicates toward molecular hydrogen formation as compared to bare silicate surfaces \citep{mennella_hydroxylated_2025}. Through our simulations we explore an additional pathway for silicate hydroxylation, either by directly capturing a hydrogen atom from colliding with a HAC nanograin or by forming hydroxylated molecular silicate fragments during the collision.

In the fragmentation regime, the high kinetic energies involved clearly facilitate the formation of a diverse range of species. Our results show that high-energy collisions between silicate and HAC grains can enable complex and largely unexplored chemical pathways that could drive a rich astrochemistry. In Table \ref{table:species_summary}, we list the most common molecular species that we find in our collisional simulations. For those species that have been observationally detected we provide the corresponding citation to the relevant literature. 

\begin{table*}
\caption{Summary of the most frequently formed bare carbon, hydrocarbon, COM, and molecular silicate species.}
\label{table:species_summary}
\centering
\renewcommand{\arraystretch}{0.9}
\begin{tabular}{l c c c c|l c c c c}
\hline\hline
\noalign{\vskip 0.5ex}
\multicolumn{10}{c}{\textbf{Summary of identified molecular species}} \\
\noalign{\vskip 0.5ex}
\hline\hline
\noalign{\vskip 0.6ex}
\makecell{Formula} & \makecell{Isomer} & \makecell{Yield} & \makecell{Velocity \\ (km s$^{-1}$)} & \makecell{Obser-\\vation} & \makecell{Formula} & \makecell{Isomer} & \makecell{Yield} & \makecell{Velocity \\ (km s$^{-1}$)} & \makecell{Obser-\\vation} \\
\noalign{\vskip 0.6ex}
\hline\hline
\noalign{\vskip 0.6ex}
\multicolumn{5}{c|}{\textbf{\textit{C$_x$}}} & \multicolumn{5}{c}{\textbf{\textit{C$_x$H$_y$}}} \\
\noalign{\vskip 0.3ex}
\hline\hline
\noalign{\vskip 0.5ex}
C$_2$         & 1 & 0.10  & 11.3            & \checkmark$^{\mathrm{(1)}}$ & C$_5$H$_2$       & 3 & 0.30  & 9.8 --11.3  & \checkmark$^{\mathrm{(23)}}$ \\
$l$-C$_3$     & 1 & 0.30  & 8.0 --11.3      & \checkmark$^{\mathrm{(2)}}$ & C$_5$H$_3$       & 3 & 0.43  & 8.9 --11.3  & x            \\
C$_4$         & 2 & 0.20  & 10.6, 11.3      & x            & C$_5$H$_4$       & 5 & 0.18  & 8.9 --11.3  & \checkmark$^{\mathrm{(24),(25)}}$ \\
C$_5$         & 2 & 0.08  & 8.0 --10.6      & \checkmark$^{\mathrm{(3)}}$ & C$_6$H$_2$       & 3 & 0.20  & 8.0, 9.8, 10.6  & \checkmark$^{\mathrm{(26),(27)}}$ \\
C$_6$         & 2 & 0.10  & 10.6, 11.3      & x            & C$_7$H$_3$       & 6 & 0.40  & 9.8 --11.3  & x            \\
\cline{6-10}
\noalign{\vskip 0.45ex}
C$_8$         & 1 & 0.10  & 8.9             & x            & \multicolumn{5}{c}{\textbf{\textit{Cyclic hydrocarbons}}} \\
\cline{1-5}
\cline{6-10}
\noalign{\vskip 0.5ex}
\multicolumn{5}{c|}{\textbf{\textit{CH$_x$}}}       & $c$-C$_3$        & 1 & 0.30  & 11.3        & x            \\
\cline{1-5}
\noalign{\vskip 0.5ex}
CH            & 1 & 0.20  & 11.3            & \checkmark$^{\mathrm{(4)}}$ & $c$-C$_3$H       & 1 & 0.15  & 8.9 --11.3  & \checkmark$^{\mathrm{(28)}}$ \\
CH$_2$        & 1 & 0.10  & 8.9, 9.8        & \checkmark$^{\mathrm{(5)}}$ & $c$-C$_3$H$_2$   & 1 & 0.14  & 8.0 --11.3  & \checkmark$^{\mathrm{(29)}}$ \\
CH$_3$        & 1 & 0.13  & 8.9, 10.6, 11.3 & \checkmark$^{\mathrm{(6)}}$ & $c$-C$_3$H$_3$   & 2 & 0.10  & 8.9 --11.3  & x            \\
\cline{1-5}
\noalign{\vskip 0.5ex}
\multicolumn{5}{c|}{\textbf{\textit{C$_x$H}}}       & C$_{11}$H$_5$    & 2 & 0.10  & 8.9, 10.6   & x            \\
\cline{1-5}
\noalign{\vskip 0.5ex}
C$_2$H        & 1 & 0.98  & 8.0 --11.3      & \checkmark$^{\mathrm{(7)}}$  & C$_{12}$H$_5$    & 1 & 0.10  & 10.6        & x            \\
$l$-C$_3$H    & 1 & 0.28  & 8.9 --11.3      & \checkmark$^{\mathrm{(8)}}$  & C$_{24}$H$_{11}$ & 1 & 0.10  & 9.8         & x            \\
C$_4$H        & 2 & 0.36  & 8.0 --11.3      & \checkmark$^{\mathrm{(9)}}$  & C$_{26}$H$_{11}$ & 1 & 0.10  & 11.3        & x            \\
C$_5$H        & 4 & 0.33  & 8.9 --11.3      & \checkmark$^{\mathrm{(10)}}$ & $c$-C$_3$HCCH    & 1 & 0.10  & 11.3        & \checkmark$^{\mathrm{(30)}}$ \\
\cline{6-10}
\noalign{\vskip 0.3ex}
C$_6$H        & 2 & 0.27  & 8.9, 10.6, 11.3 & \checkmark$^{\mathrm{(11)}}$ & \multicolumn{5}{c}{\textbf{\textit{COMs}}} \\
\cline{6-10}
\noalign{\vskip 0.5ex}
C$_7$H        & 2 & 0.20  & 10.6            & \checkmark$^{\mathrm{(12)}}$ & H$_2$C$_2$O      & 1 & 0.10  & 9.8, 11.3   & \checkmark$^{\mathrm{(31)}}$ \\
C$_8$H        & 1 & 0.08  & 8.9 --11.3      & \checkmark$^{\mathrm{(13)}}$ & HC$_3$O          & 1 & 0.10  & 8.9         & \checkmark$^{\mathrm{(32)}}$ \\
C$_9$H        & 2 & 0.10  & 10.6, 11.3      & x            & HC$_4$O          & 2 & 0.10  & 9.8, 10.6   & x            \\
C$_{10}$H     & 1 & 0.10  & 9.8             & tentative$^{\mathrm{(14)}}$ & HC$_5$O          & 1 & 0.10  & 10.6        & \checkmark$^{\mathrm{(33)}}$ \\
\cline{1-5}
\noalign{\vskip 0.3ex}
\multicolumn{5}{c|}{\textbf{\textit{C$_x$H$_y$}}}   & HC$_2$CHO        & 1 & 0.10  & 8.9, 10.6   & \checkmark$^{\mathrm{(34)}}$ \\
\cline{1-5}
\noalign{\vskip 0.5ex}
C$_2$H$_2$    & 2 & 1.34  & 6.9 --11.3      & \checkmark$^{\mathrm{(15)}}$ & C$_5$H$_2$O      & 2 & 0.10  & 8.0, 11.3   & x            \\
C$_2$H$_3$    & 1 & 0.25  & 9.8, 11.3       & \checkmark$^{\mathrm{(16)}}$ & C$_5$H$_4$O      & 1 & 0.10  & 8.0         & x            \\
$l$-C$_3$H$_2$  & 2 & 0.28  & 8.9 --11.3      & \checkmark$^{\mathrm{(17),(18)}}$ & $c$-C$_{19}$H$_6$O & 1 & 0.10 & 9.8        & x            \\
\cline{6-10}
\noalign{\vskip 0.3ex}
$l$-C$_3$H$_3$  & 2 & 0.45  & 8.9 --11.3      & \checkmark$^{\mathrm{(19)}}$ & \multicolumn{5}{c}{\textbf{\textit{Molecular Silicates}}} \\
\cline{6-10}
\noalign{\vskip 0.5ex}
C$_4$H$_2$    & 3 & 0.63  & 6.9 --11.3      & \checkmark$^{\mathrm{(20),(21)}}$ & MgSiO$_4$      & 1 & 0.20  & 9.8 --11.3  & x            \\
C$_4$H$_3$    & 3 & 0.26  & 8.0 --11.3      & x            & MgSiO$_3$        & 1 & 0.13  & 8.9 --11.3  & x            \\
C$_4$H$_4$    & 1 & 0.10  & 8.9 --11.3      & \checkmark$^{\mathrm{(22)}}$ & MgSiO$_2$        & 1 & 0.17  & 9.8 --11.3  & x            \\
\noalign{\vskip 0.3ex}
\hline\hline
\end{tabular}
\tablefoot{For each fragment, the formula, number of structural isomers, formation yield (instances per collision at productive velocities), velocity conditions for their formation, and observation status are reported. Formation yield is computed as the total number of identified instances divided by the number of simulations at the productive collision velocities (10 simulations per velocity). A yield $>1$ indicates, on average, that more than one fragment of that species is produced per collision.}

\tablebib{

(1)~\citet{C2_detection};
(2)~\citet{C3_detection};
(3)~\citet{C5_detection};
(4)~\citet{swings_ch_detection};
(5)~\citet{hollis_search_1989};
(6)~\citet{feuchtgruber_detection_2000};
(7)~\citet{tucker_ethynyl_1974};
(8)~\citet{thaddeus_astronomical_1985};
(9)~\citet{guelin_detection_1978};
(10)~\citet{cernicharo_c5h};
(11)~\citet{suzuki_detection_1986};
(12)~\citet{C7H_detection};
(13)~\citet{C8H_detection};
(14)~\citet{C10H_detection};
(15)~\citet{ridgway_circumstellar_1976};
(16)~\citet{muller_protonated_2024};
(17)~\citet{H2CCC_detection};
(18)~\citet{C3H2_detection};
(19)~\citet{CH2CCH_detection};
(20)~\citet{HC4H_C6H2_C6H6_detection};
(21)~\citet{H2C4_detection};
(22)~\citet{C4H4_detection};
(23)~\citet{H2C5_detection};
(24)~\citet{HCCCH2CCH_detection};
(25)~\citet{H2CCCHCCH_detection};
(26)~\citet{HC4H_C6H2_C6H6_detection};
(27)~\citet{HCCCHCCC_detection};
(28)~\citet{C3H_detection};
(29)~\citet{c_C3H2_detection};
(30)~\citet{C3HCCH_detection};
(31)~\citet{turner_microwave_1977};
(32)~\citet{HC3O_C5O_detection};
(33)~\citet{HC5O_detection};
(34)~\citet{HC2CHO_detection}.
}
\end{table*}

As previously discussed, hydrocarbons constitute the most populated group of species produced in our fragmentation collisions. These species are found to span a diverse range of sizes and hydrogenation levels, whose formation depends on the collision velocity range. A notable subset is the non-hydrogenated species (or pure carbon chains). We identified several of these species, ranging from C$_2$ to C$_8$, with the exception of C$_7$. The most frequent species found is $l$-C$_3$, with a formation yield of 0.30 across its productive velocity range (8.0 --11.3 km s$^{-1}$). In certain cases, we also observed the formation of non-linear isomers, such as species featuring a trimeric cyclic structure bonded to the end of a chain. A representative example is the c-C$_3C_2$ isomer of C$_5$. Detailed information regarding the structural isomers is available in the Appendix~\ref{sec:appendix_e}. We categorised the remaining hydrocarbons into three groups: CH$_n$, C$_n$H, and C$_x$H$_y$. Molecules from all three groups have been detected in space, particularly, in TMC-1 \citep{mcguire_2021_2022,xue_molecular_2025}. Our results suggest that a fraction of such molecules could originate from HAC-silicate collisions. Based on our results we may expect that hydrocarbons may also form in collisions between HAC grains, but further investigation was required to explore such processes. Among the identified members of the CH$_n$ group are methylidyne (CH), methylene (CH$_2$), and the methyl radical (CH$_3$), while methane (CH$_4$) was not found. These species have all been detected in space and are thought to play a key role in the development of hydrocarbon chemistry in the ISM through its participation in many chemical networks \citep{carbon_hydrogen_chemical_network,UMIST_2024}. The C$_n$H-type molecules, known to be radicals with linear structures, are also formed during our simulations. Specifically, we identified species ranging from the ethynyl radical (C$_2$H) to the decapentynyl radical (C$_1$$_0$H). Most molecules in this series have been previously observed, with the exception of C$_9$H,  which shows a low formation yield of 0.10 across the simulations. We also note that the observational detection of C$_1$$_0$H is tentative \citep{C10H_detection}. Generally, shorter carbon chains are produced more frequently, with C$_2$H being the most common species. Interestingly, for C$_5$H. we found four distinct structural isomers, but only one of these isomers has been thus far detected observationally. 

The third hydrocarbon group, C$_x$H$_y$, includes species with varying carbon and hydrogen counts. The most characteristic species in this group is acetylene (C$_2$H$_2$), a well known and stable molecule commonly observed in comets and the inner regions of circumstellar envelopes, where it is thought to participate in PAH formation through the HACA mechanism \citep{shukla_novel_2012,yang_hydrogenabstractionacetyleneaddition_2016}. Acetylene production in our simulations is significant, with a formation yield of 1.34 across collision velocities of 6.9 km s$^{-1}$ and above, indicating that on average more than one molecule is produced per collision event. Another astrochemically relevant species found in our simulations is vinylacetylene (C$_4$H$_4$), which is thought to play a key role in the bottom-up formation of PAHs through the complementary hydrogen abstraction-vinylacetylene addition (HAVA) mechanism \citep{HAVA_mechanism_Titan_2018}. The remainder of this group includes both cumulenes (chains of consecutive double bonds) and polyynes (chains of alternating single and triple bonds), appearing either as open-shell and closed-shell species. 

As the molecular size increases, structural isomers become more common. In some cases, collisional fragments appear as multiple observed isomers of the same chemical formula. For example, we identified the isomeric pairs propadienylidene (H$_2$CCC) and propynylidene (HCCCH), and butatrienylidene (H$_2$CCCC) and diacetylene (HCCCCH), with the latter being a closed-shell species. Another important pair involves two of the five isomers of the C$_5$H$_4$ formula, which are 1,4-pentadiyne (HCCCH$_2$CCH) and allenyl acetylene (H$_2$CCCHCCH). The latter is more reactive due to the presence of its conjugated double bonds (the allene group). Finally, a notable case is the \textit{l}-C$_7$H$_3$ radical, where two of the five identified isomers are deprotonated forms of methyltriacetylene (CH$_2$C$_6$H), a species that has already been detected in TMC-1 \citep{CH3C6H_detection}. 

An interesting case among the hydrocarbon species is the formation of cyclic molecules. Although individually their yields range from 0.10 to 0.30, their cumulative formation yield of 0.54 across all productive simulations (including the large c-C$_1$$_9$H$_6$O) suggests a non-negligible contribution to the carbon budget. These structures exhibit high chemical complexity, and their formation pathways remain poorly understood. Fragmentation from grain-grain collisions may partly provide an explanation for their formation. Interestingly, while direct formation is difficult due to the specific atom counts and higher energy inputs required compared to linear molecules, the abundant radicals produced in these collisions can facilitate the formation of cyclic species and even PAHs through radical-radical recombination. Our simulations show a high production of three-atom cyclic structures compared to larger rings. Despite the high ring strain caused by the 60° bond angles, aromaticity stabilises these three-carbon rings. This is particularly true for cyclopropenylidene (c-C$_3$H$_2$), a observationally detected molecule and the most frequent three-atom ring generated in our trajectories. Our collisions also produce larger molecules containing one to three six-atom rings substituted with linear polyynic chains. These species can be considered direct building blocks for PAH formation. These molecules tend to form at higher collisional velocities, which provide the necessary kinetic energy. However, at even higher collision velocities, the system will enter a regime of increased atomisation that reduces molecule creation. This creates a narrow velocity window with ideal conditions for their formation. It remains to be investigated whether grain-grain collisions involving only carbonaceous material, with an excess of carbon, can provide a more favourable environment for PAH formation.

Recently, COMs have gained significant attention since they represent the onset of complex organic chemistry, the chemistry in which life is rooted. The list of observationally verified COMs is continually growing  \citep{scibelli_survey_2024}. Current theories of COM formation tend to focus on gas-phase reactions or processes on grain surfaces \citep{Ceccarelli2023}. As far as we are aware, collisions between silicate and HAC grains have not been previously proposed as a COM formation route. In principle, the idea is reasonable, as silicate and HAC grains contain C, H, and O as key elements required for organic chemistry (with the exception of N). Our collision simulations identified two main types of COMs. The first type is carbonyl-containing species, such as ketene (H$_2$C$_2$O), propynal (H$_2$C$_2$CHO), and polyyne aldehydes (C$_5$H$_2$O, C$_5$H$_4$O, and c-C$_1$$_9$H$_6$O). The second COM type includes hydroxyl radicals, specifically the HC$_x$O group, where two of our three generated species have been previously observed in space. These linear molecules feature a hydrogen atom on one end and an oxygen atom on the other end. In space, these are typically thought to form through associative electron detachment reactions between carbon-chain anions (C$_x$H-) and atomic oxygen \citep{eichelberger_reactions_2007,cordiner_gas-grain_2012} or via radiative association between hydrocarbon cations and CO \citep{HC5O_detection}. Ultimately, the formation of COMs demonstrates the complex chemical diversity that can result from grain-grain collision followed by fragmentation. Overall, we find that high-velocity HAC-silicate collisions can lead to the generation of a wide variety of organic species in significant quantities. Although this initial study only considers a limited set of head-on trajectories and is limited to ultrasmall nanograins of fixed size and composition, it is remarkable that our collisions yield approximately 10\% of all currently reported organic molecules observed in space.   
Finally, we find that our collisions also lead to the formation of Mg-bearing molecular species. Here, we find several molecular silicates \citep{Bromley_mini_review}. Of particular note is the formation of the pyroxene monomer (MgSiO$_3$). This species has been produced in cluster beam experiments in both its cationic \citep{Joan_molecular_silicate_beam_2022} and anionic states \citep{Bromley_MgSiO3_anion} and could play important roles as nucleation seeds for oxygen molecules or water ice, respectively. In addition to silicate, we also find molecules based only on Mg and C atoms. Such species have recently attracted attention following the observational detection of HC$_x$Mg, and MgC$_x$N \citep{cernicharo_Mg_species}. Although we found several species with this type of chemical composition in our collisions, the isomeric structures are not the same as the currently detected species in observations.    

\section{Conclusions}

Our simulations probe a range of different collisional velocities between silicate and HAC nanograins. From our detailed physical and chemical analyses of the results of these collisions several conclusions can be drawn.

\begin{itemize}
\item We identified four collision velocity regimes. At low velocities ($\lesssim$ 1.5 km s$^{-1}$), grains bounce off each other. The bounce-and-aggregation regime ($\approx$ 1.5-3.5 km s$^{-1}$) increases the likelihood of aggregation. Intermediate velocities ($\approx$ 3.5-7.5 km s$^{-1}$) provide the optimal conditions for forming aggregated silicate-HAC nanograin species. At high velocities ($\gtrsim$ 7.5 km s$^{-1}$), fragmentation dominates, producing small molecular fragments. 

\item Although aggregated mixed nanograins can be formed through collisions, the narrow kinetic energy window for their formation indicates that this process may only occur under very specific astrophysical conditions. The long-term stability and survivability of such species under interstellar conditions is also still an open question. Nevetheless, mixed grain-grain collisions offer a potential alternative pathway for the formation of mixed grains compared to the the accretion of atoms and molecules onto existing grains \citep{jones_Themis_2013,jones_Themis_2017,ysard_themis_2024}. 

\item Several classes of molecules emerge from our HAC-silicate nanograin collisions. The most abundant species formed are hydrocarbons and CO, while silicon oxides, molecular silicates, bare carbons, COMs, and COM precursors also form. These outcomes indicate that grain-grain collisions could constitute an overlooked physical route to form a diverse range of molecular species.
\end{itemize}

Future works will consider a wider range of grain sizes, compositions, and collision angles. Such studies will help further clarify the astrochemical significance of grain-grain collisions.

\section{Data availability}

A representative trajectory for each collision velocity is publicly available on Zenodo at the following \href{https://zenodo.org/records/21099996?preview=1&token=eyJhbGciOiJIUzUxMiJ9.eyJpZCI6ImRkMmU3ZjAzLWY5MWQtNDBhMy1hNmU2LWNkNWUwZTNjMGQ2ZCIsImRhdGEiOnt9LCJyYW5kb20iOiIyNTEzYzA4ODFkZjA1MDJlMWFmZWI2NGE2NjZlMTcwZiJ9.sfDSCpyMrFpCeftwD-T8yqd3dW6bLLgtuWdNG7Gkv9sME-SSLmKPKBI54c6Nq8qdq0EimHE_0wS0YFZgbEzgSw}{link}. Each trajectory corresponds to a single NVE simulation run and includes the full atomic positions, velocities, momenta, forces, and energies at every recorded timestep.

\begin{acknowledgements}
      Part of this work was supported by funding from the European Union's Horizon 2020 research and innovation program from the European Research Council (ERC) for the project “Quantum Chemistry on Interstellar Grains” (QUANTUMGRAIN), grant agreement no 865657. MICIN (projects CNS2023-144902, PID2024-157971NB-C21 and PID2024-157971NB-C22) is also acknowledged. AR acknowledges Accademia delle Scienze di Torino for supporting the project “In silico interstellar grain-surface chemistry” and gratefully acknowledges support through the 2023 ICREA Award. Furthermore, A.K. thanks the COST Actions CA21126 NanoSpace, and CA21101 “Confined molecular systems: from a new generation of materials to the stars”(COSY) supported by COST (European Cooperation in Science and Technology) for inspiring discussions and participation in their meetings.
\end{acknowledgements}

\bibliographystyle{aa} 
\bibliography{references} 

\begin{appendix} 

\section{MACE MLIP \textit{vs} DFT comparison}
\label{sec:appendix_a}

\begin{figure}[h!]
   \centering
   \includegraphics[width=\hsize]{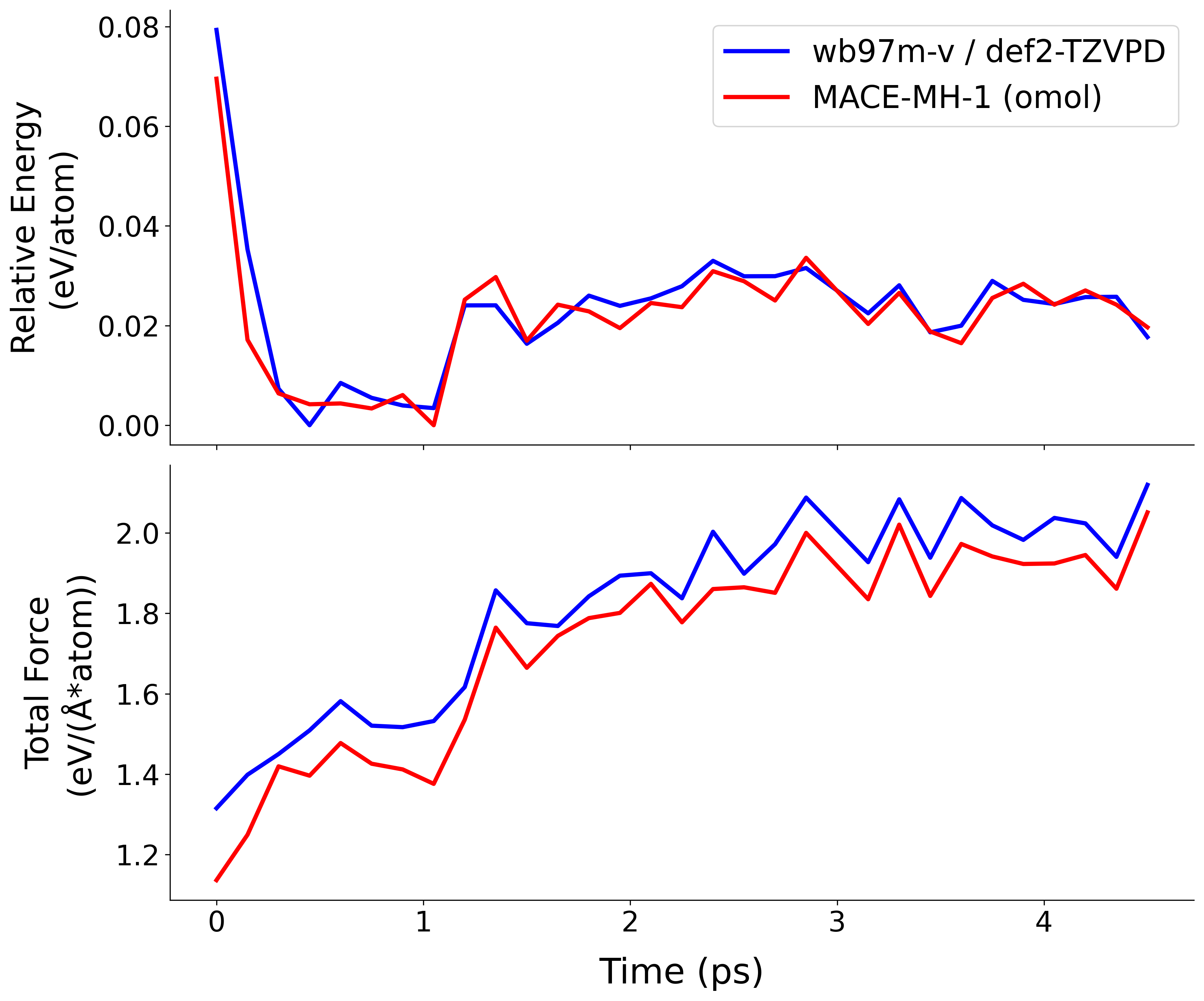}
      \caption{Comparison of relative energies (top) and absolute forces (bottom) computed per atom ($N = 186$) using the MACE-MLIP (red) and DFT with the $\omega$B97M-V functional (blue), plotted with respect to simulation time. The energies and forces were calculated along a collision trajectory generated using the MACE-MLIP.
              }
      \label{MACEvsDFT}
\end{figure}

To validate the performance of our MLIP, we extracted frames from a 2.5 km s$^{-1}$ velocity collision trajectory at intervals of 0.15 picoseconds, resulting in 30 frames spanning the collision event. Reference energies and forces were computed using density functional theory (DFT) calculations using the $\omega$B97M-V functional \citep{wB97M_V_functional} and a def2-TZVPD basis set, as implemented in ORCA 6.0 \citep{ORCA_6_0}. This level of theory was chosen to maintain consistency with the fine-tuned MLIP training dataset. In  Figure~\ref{MACEvsDFT}, we show the relative energies per atom (referenced to the lowest energy point in each case) and absolute forces per atom for the collision calculated using our MACE-MLIP compared to the DFT reference values. Averaged across all frames in the trajectory, the relative energies from our MLIP simulation show a small mean absolute difference of $3.26 \times 10^{-3}$ eV/atom compared to the DFT reference, confirming good agreement. For the forces, the mean absolute difference is $9.01 \times 10^{-2}$ eV/(\AA$\cdot$atom). Here, the MLIP systematically underestimates these forces compared to the DFT values, which is consistent with expectations \citep{deng_systematic_2025}. As noted in the methodology, MLIPs tend to 'soften' the potential energy surface and the resulting forces, when trained on datasets mainly containing energy minimised structures. This small systematic underestimation of forces in the high-compression, short-range repulsive regime will likely somewhat dampen the peak shock pressure in the collisions. This effect may lead to a slight attenuation of the thermodynamic scale of our results, where our trends with respect to increasing collisional velocity are maintained but the predicted onset for each progressive phase occurs as a slightly overestimated velocity. This would mean that our fragmentation velocities should be taken as conservative upper bounds. Importantly, as the MLIP is well-trained on the chemistry, the chemical pathways explored in the collisions remain physically sound. As such, this issue does not affect the reliability of the MLIP to accurately represent the species produced in collisional events.

\FloatBarrier

\section{Timestep selection}
\label{sec:appendix_b}

\begin{figure}[h!]
   \centering
   \includegraphics[width=0.95\hsize]{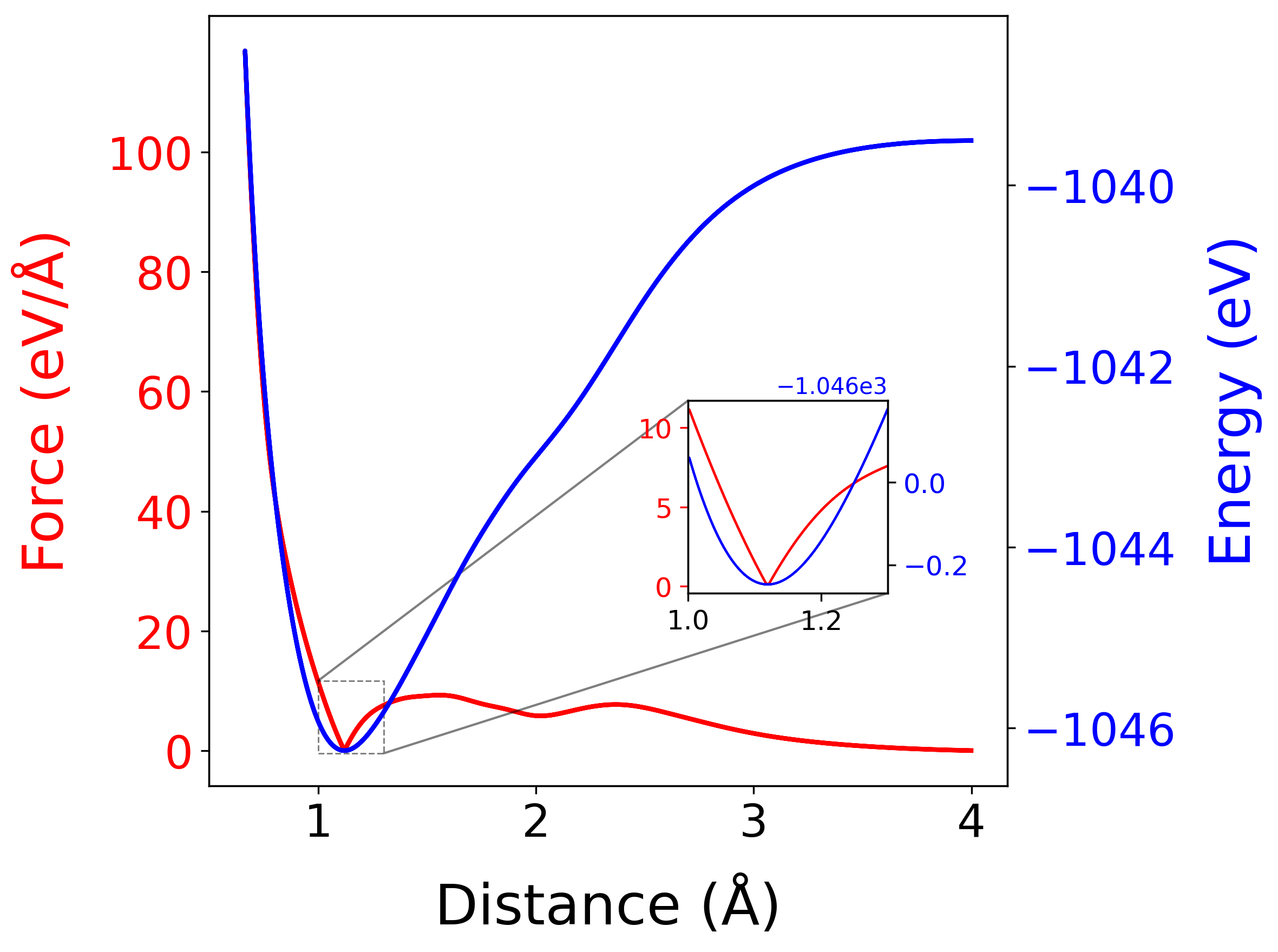}
      \caption{Force (red line) and potential energy (blue line) as a function of C-H interatomic distance, calculated using the MACE MLIP with a 0.01 fs timestep. The region between 1-1.20 Å, where the force transitions between attractive and repulsive, is highlighted.}
      \label{Timestep}
\end{figure}

\begin{figure}[h!]
   \centering
   \includegraphics[width=0.95\hsize]{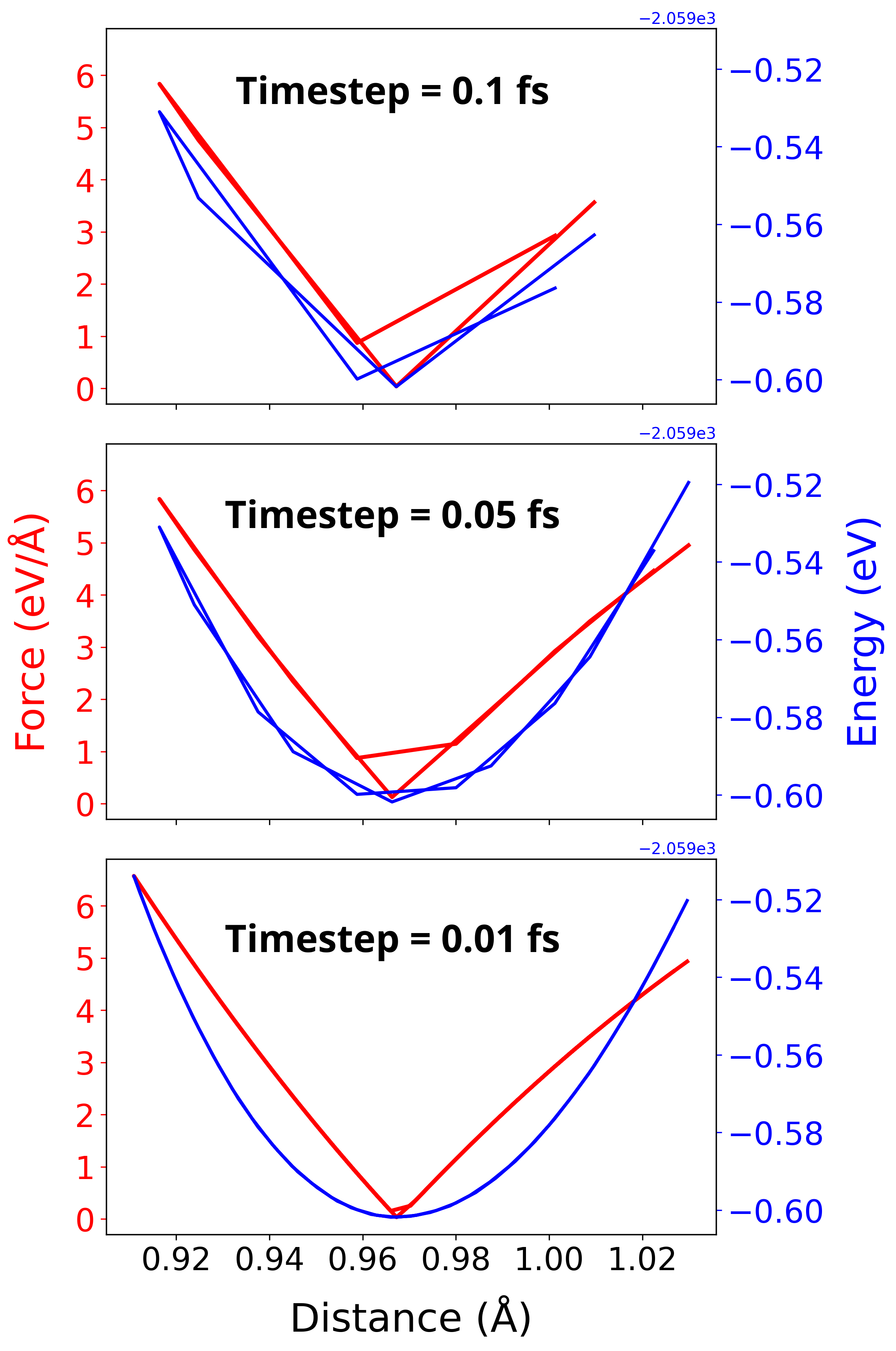}
      \caption{Force (red line) and potential energy (blue line) profiles as a function of the O-H interatomic distance calculated using the MACE MLIP for three timestep values: 0.1 fs (top), 0.5 fs (medium), and 0.01 fs (bottom). The profiles highlight the region, where the force transition occurs.}
      \label{Timestep_CO_pair}
\end{figure}

As previously discussed, the timestep is a critical parameter in these simulations, especially given the high velocities we have specified. A small timestep can result in significantly longer computational times, particularly when using a relatively complex interatomic potential. On the contrary, a large timestep can introduce artefacts into the simulation. Specifically, in the force profile between a pair of atoms, a timestep that is too large can cause unexpected transitions from attractive to repulsive forces, leading to misleading results. A suitably chosen timestep should ensure a smooth force trajectory while being as computationally efficient as possible. 

To determine a suitable timestep for our collisions, we conducted a test in which pairs of atoms were set to collide, and we monitored the forces and potential energy throughout the simulations. The simulations used the NVE ensemble and an initial total velocity of 7.5 km s$^{-1}$, representative of our typical collision velocities. We evaluated three candidate timesteps: 0.1, 0.5, and 0.01 fs. Figure~\ref{Timestep} shows the potential energy and force profiles for a collision between a carbon atom and a hydrogen atom as a function of the C-H distance. When focusing on the region where the force transitions from attractive to repulsive and vice versa, the effect of the timestep magnitude is most evident. Specifically, as shown in Figure~\ref{Timestep_CO_pair} for a collision between an oxygen atom and a hydrogen atom, the force exhibits discontinuous jumps when using 0.1 and 0.05~fs timesteps. This artefact was consistently observed for all atom pairs, albeit with varying magnitudes. Among the tested timesteps, 0.01 fs was identified as the largest timestep that maintained an acceptable physical accuracy for collisions at these velocities was thus used in all simulations.

\FloatBarrier

\section{Internal temperatures after the collisions}
\label{sec:appendix_c}

To characterise the energetic state of the collision products and assess the potential role of electronic effects, we computed the internal temperature of the largest Si-based and C-based species at the end of each simulation. For each fragment the centre of mass velocity and rotation were subtracted, retaining only the internal vibrational contribution to the kinetic energy. The internal temperature was then obtained via the equipartition theorem,

\begin{equation}
    T_{\mathrm{int}} = \frac{2\,E_{\mathrm{kin}}^{\mathrm{int}}}
    {(3N - 6)\,k_{\mathrm{B}}},
\end{equation}

\begin{table}[h!]
\centering
\caption{Mean internal kinetic temperatures of the largest Si-based and C-based collision species.}
\label{tab:temperatures}
\begin{tabular}{ccc|cc}
\hline
\hline
\noalign{\vskip 0.5ex}
$v_{\mathrm{coll}}$ & \multicolumn{2}{c|}{$T_{\mathrm{int}}$ (Si-based) (K)} 
& \multicolumn{2}{c}{$T_{\mathrm{int}}$ (C-based) (K)} \\
\cline{2-3} \cline{4-5}
\noalign{\vskip 0.5ex}
(km s$^{-1}$) & Mean & $\sigma$ & Mean & $\sigma$ \\
\hline
\noalign{\vskip 0.5ex}
0.8  & 284.6  & 316.3  & 859.8  & 72.8  \\
1.8  & 595.6  & 184.6  & 826.7  & 76.5  \\
2.5  & 929.2  & 139.2  & 989.5  & 105.6 \\
4.0  & 1532.2 & 88.9   & 1532.2 & 88.9  \\
5.6  & 2418.2 & 187.2  & 2418.2 & 187.2 \\
6.9  & 3108.1 & 217.0  & 3108.1 & 217.0 \\
8.0  & 3443.7 & 169.8  & 3443.7 & 169.8 \\
8.9  & 3711.2 & 856.6  & 3555.2 & 303.8 \\
9.8  & 3902.1 & 676.5  & 3532.9 & 472.1 \\
10.5 & 4411.2 & 657.3  & 3861.9 & 772.3 \\
11.3 & 5247.6 & 1043.3 & 3749.5 & 747.1 \\
\hline
\hline
\end{tabular}
\tablefoot{Values are averaged over ten independent runs. Uncertainties ($\sigma$) represent one standard deviation.}
\end{table}

\noindent where $N$ is the number of atoms in the fragment, $3N-6$ are 
the internal degrees of freedom (excluding overall translation and 
rotation), and $k_{\mathrm{B}}$ is the Boltzmann constant. The mean internal temperatures averaged over the species from the ten independent runs per velocity are reported in Table~\ref{tab:temperatures}. At intermediate velocities (4.0 - 8.0 km s$^{-1}$), the Si-based and C-based nanograins aggregate, and the reported temperature therefore corresponds to the combined structure.

\noindent The deposited energy is distributed over a large number of vibrational modes, keeping the excitation per atom comparatively low even in the fragmentation regime. In all cases, the reported internal temperatures correspond to energies well below those required for electronic excitation, supporting the conclusion that hot-atom electronic effects are unlikely to play a significant role in the collision outcomes considered here. Even for the smallest and most abundant molecular fragments (e.g C$_2$H$_2$) the internal energies remain well below their respective bond dissociation energies, indicating that unimolecular decay is not expected on the timescale of our simulations.

\FloatBarrier

\section{Overall table}
\label{sec:appendix_d}

Table \ref{table:Summary_energies_1} summarise the average outcomes of our simulated collisions with respect to the ten independent runs (i.e. where for each run the nanograins are randomly rotated) for each velocity. The data include the average total velocity of the species, the average number of produced species, and the average number of species containing carbon and silicon. Additionally, it provides the average number of specific molecular species such as CO, CO$_2$, H$_2$O, H$_2$, OH, HCO, Hydrocarbons (HC), Si$_y$O$_x$, COMs, mixed clusters, molecular silicates, and bare carbons. It concludes with the kinetic energy (KE) per atom for each velocity before and after the collision. This table serves as a comprehensive summary of the chemical diversity generated by our mixed nanograin collisions across the range of velocities studied.

\begin{table}[h!]
\caption{Occurrence of selected species produced in the collisions with respect to collisional velocity averaged over the ten independent collisions for each velocity.}             
\label{table:Summary_energies_1}      
\centering
\renewcommand{\arraystretch}{1.30}
\resizebox{\columnwidth}{!}{%
\begin{tabular}{p{5cm} *{6}{c}}
\hline
\hline
\noalign{\vskip 0.5ex}
Initial KE/\textit{N} (eV) & 0.01 & 0.05 & 0.1 & 0.25 & 0.5 & 0.75 \\
\hline
\hline
\noalign{\vskip 0.5ex}
Total velocity (km s$^{-1}$) & 0.8 & 1.8 & 2.5 & 4.0 & 5.6 & 6.9 \\
Number of species & 2.0 & 1.6 & 1.6 & 1.0 & 1.2 & 2.8 \\
Carbon containing species & 1.0 & 1.0 & 1.0 & 1.0 & 1.2 & 2.3 \\
Silicon containing species & 1.0 & 1.0 & 1.0 & 1.0 & 1.0 & 1.2 \\
CO & 0.0 & 0.0 & 0.0 & 0.0 & 0.1 & 0.8 \\
CO$_2$ & 0.0 & 0.0 & 0.0 & 0.0 & 0.0 & 0.0 \\
H$_2$O & 0.0 & 0.0 & 0.0 & 0.0 & 0.0 & 0.0 \\
H$_2$ & 0.0 & 0.0 & 0.0 & 0.0 & 0.0 & 0.0 \\
\noalign{\vskip 0.3ex}
\hline
\hline
\end{tabular}%
}
\end{table}

\begin{table}[h!]
\addtocounter{table}{-1}
\caption{continued.}             
\label{table:Summary_energies_2}      
\centering
\renewcommand{\arraystretch}{1.30}
\resizebox{\columnwidth}{!}{%
\begin{tabular}{p{5cm} *{6}{c}}
\hline
\hline
\noalign{\vskip 0.5ex}
Initial KE/\textit{$N$} (eV) & 0.01 & 0.05 & 0.1 & 0.25 & 0.5 & 0.75 \\
\hline
\hline
\noalign{\vskip 0.5ex}
OH & 0.0 & 0.0 & 0.0 & 0.0 & 0.0 & 0.0 \\
HCO & 0.0 & 0.0 & 0.0 & 0.0 & 0.0 & 0.1 \\
HC & 1.0 & 0.6 & 0.6 & 0.0 & 0.0 & 0.3 \\
Si$_y$O$_x$ & 0.0 & 0.0 & 0.0 & 0.0 & 0.0 & 0.1 \\
COMs & 0.0 & 0.0 & 0.0 & 0.0 & 0.0 & 0.0 \\
Mixed clusters & 0.0 & 0.4 & 0.4 & 1.0 & 1.0 & 1.0 \\
Molecular silicates & 0.0 & 0.0 & 0.0 & 0.0 & 0.0 & 0.0 \\
Pure carbon-based species & 0.0 & 0.0 & 0.0 & 0.0 & 0.0 & 0.0 \\
KE/\textit{$N$} before contact (eV) & 0.067 & 0.088 & 0.138 & 0.288 & 0.538 & 0.788 \\
KE/\textit{$N$} after contact (eV) & 0.090 & 0.104 & 0.130 & 0.194 & 0.311 & 0.395 \\
\hline
\hline
\noalign{\vskip 0.5ex}
Initial KE/\textit{$N$} (eV) & 1.0 & 1.25 & 1.5 & 1.75 & 2.0 & \\
\hline
\hline
\noalign{\vskip 0.5ex}
Total velocity (km s$^{-1}$) & 8.0 & 8.9 & 9.8 & 10.6 & 11.3 & \\
Number of species & 7.7 & 16.3 & 24.6 & 30.9 & 31.5 & \\
Carbon containing species & 6.7 & 13.7 & 19.9 & 23.8 & 23.3 & \\
Silicon containing species & 1.3 & 3.2 & 3.8 & 5.5 & 5.6 & \\
CO & 2.0 & 4.8 & 5.9 & 5.8 & 6.5 & \\
CO$_2$ & 0.2 & 0.3 & 0.1 & 0.5 & 0.2 & \\
H$_2$O & 0.0 & 0.1 & 0.1 & 0.1 & 0.0 & \\
H$_2$ & 0.0 & 0.1 & 0.3 & 0.2 & 0.1 & \\
OH & 0.0 & 0.0 & 0.0 & 0.0 & 0.1 & \\
HCO & 0.1 & 0.0 & 0.3 & 0.4 & 0.4 & \\
HC & 2.6 & 5.6 & 9.3 & 12.4 & 11.7 & \\
Si$_y$O$_x$ & 0.2 & 0.6 & 1.3 & 2.3 & 1.9 & \\
COMs & 0.0 & 0.3 & 0.9 & 1.0 & 0.9 & \\
Mixed clusters & 1.0 & 1.6 & 1.9 & 1.7 & 1.2 & \\
Molecular silicates & 0.1 & 0.8 & 0.6 & 1.0 & 1.7 & \\
Pure carbon-based species & 0.1 & 0.2 & 0.4 & 0.8 & 1.4 & \\
KE/\textit{$N$} before contact (eV) & 1.038 & 1.288 & 1.538 & 1.788 & 2.038 & \\
KE/\textit{$N$} after contact (eV) & 0.457 & 0.539 & 0.637 & 0.782 & 0.930 & \\
\noalign{\vskip 0.3ex}
\hline
\hline
\end{tabular}%
}
\tablefoot{Energies are computed per atom ($N = 186$). The Induced KE/\textit{$N$} represents the translational kinetic energy applied to the system due to the initial velocities given to the grains.}
\end{table}

\FloatBarrier

\section{Isomer classification}
\label{sec:appendix_e}

Table \ref{table:isomer_breakdown_1} provides a detailed breakdown of the organic structural isomers identified in our simulated collisions, along with their geometric structures, yield of formation, the range of collision velocities at which they were formed, and their current observational status.

\begin{table}[h!]
\caption{Isomeric breakdown of organic molecular species identified in collision simulations.}
\label{table:isomer_breakdown_1}
\centering
\renewcommand{\arraystretch}{1.15}
\footnotesize
\setlength{\tabcolsep}{4pt} 
\begin{tabular}{l l c c c}
\hline\hline
\noalign{\vskip 0.5ex}
\textbf{Formula} & \textbf{Geometric formula} & \textbf{Yield} & \textbf{Velocity} (km s$^{-1}$) & \textbf{Obs.} \\
\noalign{\vskip 0.5ex}
\hline\hline
\noalign{\vskip 0.5ex}
C$_4$          & CCCC                            & 0.10  & 10.6        & x          \\
C$_4$          & $c$-C$_3$C                      & 0.30  & 11.3        & x          \\
C$_5$          & CCCCC                           & 0.10  & 10.6        & \checkmark \\
C$_5$          & $c$-C$_3$CC                     & 0.10  & 8.0, 9.8    & x          \\
C$_6$          & CCCCCC                          & 0.10  & 11.3        & x          \\
C$_6$          & $c$-C$_3$CCC                    & 0.10  & 10.6        & x          \\
C$_4$H         & CCCCH                           & 0.32  & 8.0--11.3   & \checkmark \\
C$_4$H         & $c$-C$_3$HC                     & 0.10  & 9.8, 11.3   & x          \\
\noalign{\vskip 0.3ex}
\hline\hline
\end{tabular}
\end{table}

\begin{table}[h!] 
\label{table:isomer_breakdown_2}
\centering
\renewcommand{\arraystretch}{1.15}
\footnotesize
\setlength{\tabcolsep}{4pt}
\begin{tabular}{l l c c c}
\hline\hline
\noalign{\vskip 0.5ex}
\textbf{Formula} & \textbf{Geometric formula} & \textbf{Yield} & \textbf{Velocity} (km s$^{-1}$) & \textbf{Obs.} \\
\noalign{\vskip 0.5ex}
\hline\hline
\noalign{\vskip 0.5ex}
C$_5$H         & CCCCCH                          & 0.17  & 8.9--10.6   & \checkmark \\
C$_5$H         & $c$-C$_3$CCH                    & 0.10  & 8.9--11.3   & x          \\
C$_5$H         & $c$-C$_3$HCC                    & 0.30  & 11.3        & x          \\
C$_5$H         & CCCHCC                          & 0.10  & 8.9         & x          \\
C$_6$H         & CCCCCCH                         & 0.15  & 8.9--11.3   & \checkmark \\
C$_6$H         & $c$-C$_3$HCCC                   & 0.10  & 10.6, 11.3  & x          \\
C$_7$H         & CCCCCCCH                        & 0.10  & 10.6        & \checkmark \\
C$_7$H         & CCCHCCCC                        & 0.10  & 10.6        & x          \\
C$_9$H         & CCCCCCCCCH                      & 0.10  & 10.6        & x          \\
C$_9$H         & $c$-C$_3$CCCCCCH                & 0.10  & 11.3        & x          \\
C$_2$H$_2$     & HCCH                            & 1.22  & 6.9--11.3   & \checkmark \\
C$_2$H$_2$     & H$_2$CC                         & 0.18  & 8.9--11.3   & x          \\
$l$-C$_3$H$_2$ & HCCCH                           & 0.23  & 8.9--11.3   & \checkmark \\
$l$-C$_3$H$_2$ & H$_2$CCC                        & 0.20  & 11.3        & \checkmark \\
$l$-C$_3$H$_3$ & H$_2$CCCH                       & 0.40  & 8.9--11.3   & \checkmark \\
$l$-C$_3$H$_3$ & H$_3$CCC                        & 0.10  & 10.6, 11.3  & x          \\
C$_4$H$_2$     & HCCCCH                          & 0.57  & 6.9--11.3   & \checkmark \\
C$_4$H$_2$     & H$_2$CCCC                       & 0.15  & 9.8, 10.6   & \checkmark \\
C$_4$H$_2$     & HCCCHC                          & 0.10  & 10.6        & x          \\
C$_4$H$_3$     & H$_2$CCCCH                      & 0.18  & 8.0--11.3   & x          \\
C$_4$H$_3$     & H$_2$CCHCC                      & 0.10  & 8.0, 11.3   & x          \\
C$_4$H$_3$     & $c$-C$_3$HCH$_2$                & 0.10  & 9.8, 11.3   & x          \\
C$_5$H$_2$     & HCCCCCH                         & 0.20  & 9.8--11.3   & \checkmark \\
C$_5$H$_2$     & H$_2$CCCCC                      & 0.10  & 10.6, 11.3  & x          \\
C$_5$H$_2$     & HCCCHCC                         & 0.10  & 9.8         & x          \\
C$_5$H$_3$     & H$_2$CCCCCH                     & 0.20  & 8.9--11.3   & x          \\
C$_5$H$_3$     & HCCCHCCH                        & 0.20  & 8.9--11.3   & x          \\
C$_5$H$_3$     & $c$-C$_3$H$_2$CCH               & 0.10  & 9.8         & x          \\
C$_5$H$_4$     & H$_2$CCCHCCH                    & 0.15  & 9.8, 10.6   & \checkmark \\
C$_5$H$_4$     & HCCC[H$_2$]CCH                  & 0.10  & 8.9         & \checkmark \\
C$_5$H$_4$     & H$_2$CCHCHCC                    & 0.10  & 10.6        & x          \\
C$_5$H$_4$     & HCCCCHCH$_2$                    & 0.10  & 9.8         & x          \\
C$_5$H$_4$     & $c$-C$_3$HCHCH$_2$              & 0.10  & 11.3        & x          \\
C$_6$H$_2$     & HCCCCCCH                        & 0.10  & 8.0--10.6   & \checkmark \\
C$_6$H$_2$     & HCCCHCCC                        & 0.10  & 10.6        & \checkmark \\
C$_6$H$_2$     & $c$-C$_3$CCCH                   & 0.10  & 9.8         & x          \\
C$_7$H$_3$     & H$_2$CCCCCCCH                   & 0.17  & 9.8--11.3   & x          \\
C$_7$H$_3$     & HCCCHCCHCC                      & 0.10  & 10.6        & x          \\
C$_7$H$_3$     & HCCCHCCCCH                      & 0.15  & 9.8, 10.6   & x          \\
C$_7$H$_3$     & H$_3$CCCCCCC                    & 0.10  & 9.8         & x          \\
C$_7$H$_3$     & $c$-C$_3$H$_2$CHCCC             & 0.10  & 9.8         & x          \\
C$_7$H$_3$     & $c$-C$_3$HCHCCCH                & 0.10  & 11.3        & x          \\
$c$-C$_3$H$_3$ & $c$-CHCH$_2$C                   & 0.10  & 8.9, 11.3   & x          \\
$c$-C$_3$H$_3$ & $c$-CHCHCH                      & 0.10  & 8.9, 9.8    & x          \\
C$_{11}$H$_5$  & $c$-C$_5$H$_4$CCCCCCH           & 0.10  & 8.9         & x          \\
C$_{11}$H$_5$  & $c$-C$_5$H$_4$(CCH)($c$-C$_3$C) & 0.10  & 10.6        & x          \\
HC$_4$O        & HCCCCO                          & 0.10  & 9.8         & x          \\
HC$_4$O        & CCCHCO                          & 0.10  & 10.6        & x          \\
C$_5$H$_2$O    & HCCCCCHO                        & 0.10  & 8.0         & x          \\
C$_5$H$_2$O    & HCCCHCCO                        & 0.10  & 11.3        & x          \\
\noalign{\vskip 0.3ex}
\hline\hline
\end{tabular}
\tablefoot{For each structural isomer, the geometric formula, formation yield (instances per collision at productive collision velocities), velocity conditions for their formation, and observation status are reported.}
\end{table}

\end{appendix}
\end{document}